\documentclass[journal,10pt,a4paper,twocolumn,twoside]{IEEEtran}

\IEEEoverridecommandlockouts

\usepackage[active]{srcltx} 
\usepackage{graphicx}
\usepackage{balance}
\usepackage{cite}
\usepackage{amssymb}
\usepackage[cmex10]{amsmath}
\usepackage{amsfonts}
\usepackage{amsthm}

\usepackage{eurosym}
\usepackage{cuted} 

\usepackage{booktabs}
\usepackage{hyperref}

\usepackage{multirow}

\usepackage{algorithm}
\usepackage{algorithmicx}
\usepackage{algpseudocode}

\usepackage{subcaption}

\usepackage{braket}
\usepackage{tikz}
\usetikzlibrary{quantikz}
\usepackage{adjustbox}
\usepackage{easyReview}

\usepackage{rotating}

\usepackage{placeins}

\usepackage{dblfloatfix}

\begin{document}

\newcommand{\meterCom}{\scalebox{.5}{\begin{tikzcd} \meter{} \end{tikzcd}}}
\newcommand{\cwCom}{\scalebox{.5}{\begin{tikzcd} \pgfmatrixnextcell \cw \end{tikzcd}}}

\newcommand{\eqdef}{\stackrel{\triangle}{=}}

\newtheorem{theo}{Theorem}
\newtheorem{theor}{Theorem}
\newtheorem{cor}{Corollary}
\newtheorem{lem}{Lemma}
\newtheorem{prop}{Proposition}
\newtheorem{ins}{Insight}
\newtheorem{remark}{Remark}

\theoremstyle{remark}

\theoremstyle{definition}
\newtheorem{defin}{Definition}
\newtheorem{ass}{Assumption}
\newtheorem{rem}{Remark}

\renewcommand{\qed}{$\blacksquare$}

\renewcommand{\algorithmiccomment}[1]{// #1}

\title{Beyond Shannon Limits: Quantum Communications through Quantum Paths}

\author{Marcello~Caleffi,~\IEEEmembership{Senior~Member,~IEEE}, Kyrylo~Simonov, Angela~Sara~Cacciapuoti$^*$,~\IEEEmembership{Senior~Member,~IEEE}	\thanks{A.S. Cacciapuoti and M. Caleffi are with the the \href{www.quantuminternet.it}{www.QuantumInternet.it} research group, \textit{FLY: Future Communications Laboratory}, at the Department of Electrical Engineering and Information Technology (DIETI), University of Naples Federico II, Naples, 80125 Italy. K. Simonov is independent researcher. E-mail:  \href{mailto:angelasara.cacciapuoti@unina.it}{angelasara.cacciapuoti@unina.it}, \href{mailto:kyrylo.simonov@univie.ac.at}{kyrylo.simonov@univie.ac.at}, \href{mailto:marcello.caleffi@unina.it}{marcello.caleffi@unina.it}. Web: \href{http://www.quantuminternet.it}{www.quantuminternet.it}.}
	\thanks{A.S. Cacciapuoti and M. Caleffi are also with the Laboratorio Nazionale di Comunicazioni Multimediali, National Inter-University Consortium for Telecommunications (CNIT), Naples, 80126, Italy.}
	\thanks{This work was partially supported by the project ``\textit{Towards the Quantum Internet: A Multidisciplinary Effort}'', University of Naples Federico II, Italy.}
	\thanks{$^*$Corresponding author.}
}
\date{today}

\maketitle

\begin{abstract}
A crucial step towards the 6th generation (6G) of networks would be a shift in communication paradigm beyond the limits of Shannon's theory. In both classical and quantum Shannon's information theory, communication channels are generally assumed to combine through \textit{classical trajectories}, so that the associated network path traversed by the information carrier is well-defined. Counter-intuitively, quantum mechanics enables a quantum information carrier to propagate through a \textit{quantum path}, i.e., through a path such that the causal order of the constituting communications channels becomes indefinite. Quantum paths exhibit astonishing features, such as providing non-null capacity even when no information can be sent through any classical path. In this paper, we study the quantum capacity achievable via a quantum path and establish upper and the lower bounds for it. Our findings reveal the substantial advantage achievable with a quantum path over any classical placements of communications channels in terms of ultimate achievable communication rates. Furthermore, we identify the region where a quantum path incontrovertibly outperforms the amount of transmissible information beyond the limits of conventional quantum Shannon's theory, and we quantify this advantage over classical paths through a conservative estimate. \end{abstract}

\begin{IEEEkeywords}
Quantum Capacity, Quantum Path, Quantum Trajectory, Quantum Switch, Entanglement, Causal Order, Classical Path, Quantum Internet.
\end{IEEEkeywords}

\section{Introduction}
\label{sec:1}

\subsection{Motivation}

     \begin{figure}[t]
        \centering
		\includegraphics[width=1\columnwidth]{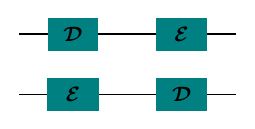}
		\caption{Information carrier traversing channel $\mathcal{E}$ after channel $\mathcal{D}$ or vice versa, i.e., through a \textit{classical path} implementing the well-defined causal order $\mathcal{D \rightarrow E}$ or $\mathcal{E \rightarrow D}$, respectively.}
		\label{Fig:1}
  \hrulefill
     \end{figure}
     
	\begin{figure}[t]
        \centering
		\includegraphics[width=1\columnwidth]{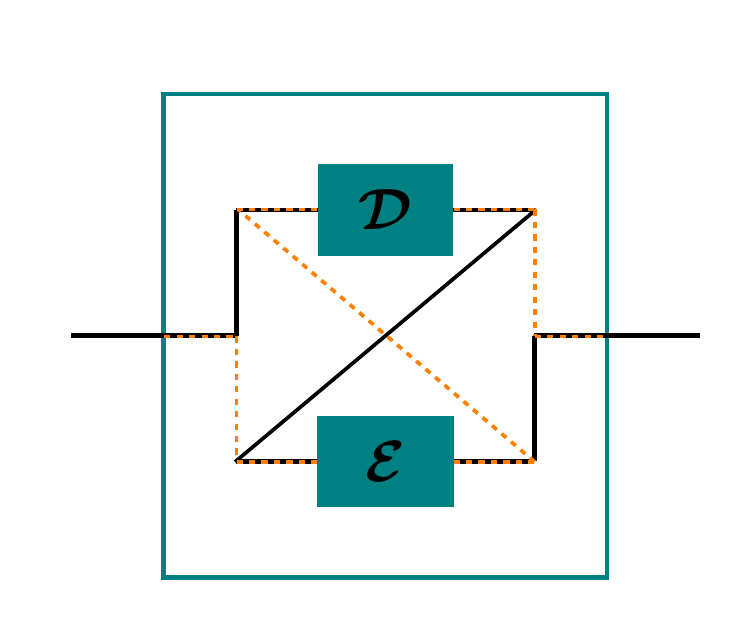}
		\caption{Information carrier traversing channels $\mathcal{D}$ and $\mathcal{E}$ in a superposition of the two alternative causal orders $\mathcal{D \rightarrow E}$ and $\mathcal{E \rightarrow D}$, i.e., through a \textit{quantum path}.}
		\label{Fig:3}
  \hrulefill
      \end{figure}

Despite a significant advancement in communications after recent introduction of the 5th generation (5G) networks, the fast-growing needs of exchanging ever bigger amount of data via more effective new services trigger development of the 6th generation (6G) networks. A crucial problem is reaching ever higher communication rates, which, however, are limited by the Shannon's information theory. Indeed, the fundamental assumption underlying Shannon's theory was to model both the information carriers and the communication channels as classical entities, obeying the law of classical physics \cite{KouCacSim-22,ChiKri-19}.

During the last five decades, scientists have been working on extending Shannon's theory to the quantum domain\footnote{For a comprehensive overview of quantum communication field, the interested reader might refer to the following works \cite{KozWehVan-22,KouCacSim-22,CacCalVanHan-19,CacCalTaf-18,GyIm18,PirEisWee15}.} -- an area of study referred to as \textit{quantum Shannon's theory} \cite{Wil-13,Pre-18} -- by modeling the information carriers as quantum systems and by exploiting the unconventional phenomena -- superposition and entanglement -- arising from such a modeling. The reason behind such an extension lies in the capability of quantum mechanics to enable applications in communication networks with no-counterpart in classical networks, as pointed out in \cite{IllCalMan-22,CacIllKou-22,WanRahLi-22}.

\begin{figure*}[t!]
	\centering
	\includegraphics[width=1\linewidth]{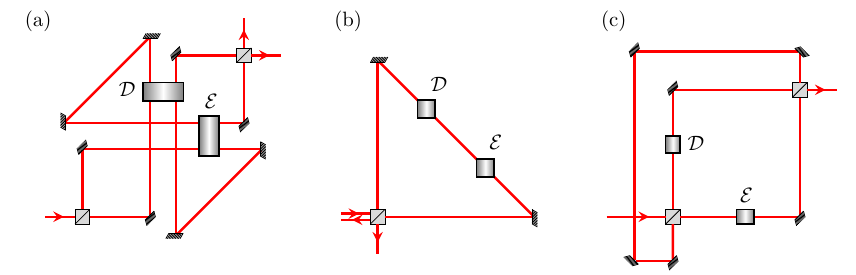}
	\caption{Schematic diagram of some of possible architectures of the photonic quantum SWITCH. (a) An implementation via a Mach-Zehnder geometry, where the target qubit is encoded in polarization of the photon, while the control qubit is mapped into its path degree of freedom using the first beam splitter and coherently recombining the paths $\mathcal{D} \rightarrow \mathcal{E}$ and $\mathcal{E} \rightarrow \mathcal{D}$ at the second beam splitter~\cite{ProMoqAra-15,RubRozFei-17,GuoHuHou-18,Pro-19,RubTozMas-22}. (b) An implementation via a Sagnac geometry, where the target qubit is encoded in polarization of the photon (as in (a)), whereas a single beam splitter introduces the path degree of freedom as control and completes superposition of causal orders of $\mathcal{D}$ and $\mathcal{E}$~\cite{StrSchPet-22}. (c) An implementation via a geometry, where the target qubit is encoded in the path degree of freedom of the photon, while the role of control qubit is played by its polarization~\cite{GosGiaKew-18, GosRomWhi-18}.}
	\label{Fig:4}
	\hrulefill
\end{figure*}

But, in both classical and quantum Shannon's information theory, communication channels are generally assumed to be placed in a classical setting, such that the associated network path traversed by the information carrier is well-defined. As instance, with reference to Fig.~\ref{Fig:1}, when a message $m$ is intended to propagate from the sender through two communication channels -- say channels $\mathcal{D}$ and $\mathcal{E}$ -- to reach the receiver, either channel $\mathcal{E}$ is crossed after channel $\mathcal{D}$ or vice versa. In both cases, the causal order of the channels traversed by the message is well-defined, i.e., either $\mathcal{D \rightarrow E}$ or $\mathcal{E \rightarrow D}$.

\begin{table*}[h!]
    \centering
       \begin{tabular}{| p{0.15\textwidth} |   p{0.60\textwidth}|p{0.1\textwidth}|}
		\toprule
		    \textbf{Symbol} & \textbf{Definition} &\textbf{Appearance} \\
        \midrule
            $\mathcal{N}(\cdot)$ & A quantum channel &  Section~\ref{sec:2}\\
        \midrule
		    $\mathbf{H}_A$ &  Hilbert space of a system $A$ &  Section~\ref{sec:2}\\
        \midrule
            $\ket{\psi_A}$ & Pure state of a system $A$&  Section~\ref{sec:2}\\ 
        \midrule
            $\rho_A$ & Density operator of a system $A$ &  Section~\ref{sec:2}\\
        \midrule
            $\mathcal{L}(\mathbf{H}_A)$  & The set of density operators on the Hilbert space $\mathbf{H}_A$ &  Section~\ref{sec:2}\\
        \midrule
            $\mathcal{I}_A$  & Identity channel for a system $A$ &  Section~\ref{sec:2}\\
        \midrule
            $\otimes$  & Tensor product &  Section~\ref{sec:2} \\
        \midrule
            $\ket{\Phi^+}$ & Bipartite maximally entangled state &  Section~\ref{sec:2}\\ 
        \midrule
            $\rho_{\mathcal{N}}$  & The Choi matrix of the channel $\mathcal{N}$ & Section~\ref{sec:2}\\
        \midrule
            $I_c(\rho,\mathcal{N})$   & The coherent information of the channel $\mathcal{N}$ with respect to the input state $\rho$ &  Section~\ref{sec:2} \\
        \midrule
            $S(\rho)$  & The von Neumann entropy of the density operator $\rho$ & Section~\ref{sec:2}\\
        \midrule
            $I_c(\mathcal{N})$   & The coherent information of the channel $\mathcal{N}$ &  Section~\ref{sec:2} \\
        \midrule
            $Q(\cdot)$ & The quantum capacity of a quantum channel & Section~\ref{sec:2}\\
		\midrule
            $S(\rho||\sigma)$  & Quantum relative entropy of a state $\rho$ with respect to a state $\sigma$ &  Section~\ref{sec:2}\\
        \midrule
            $E_R(\rho)$  & Quantum entropy of entanglement of a state $\rho$ &  Section~\ref{sec:2}\\
        \midrule
            $\mathcal{N}_{\text{C}}(\cdot)(\cdot)$ & Overall channel from a classical path &  Section~\ref{sec:2}\\
        \midrule
            $Q_{\text{C}}^{\text{UB}}$ & Upper bound on quantum capacity of the classical path & Section~\ref{sec:2}\\
		\midrule
            $A^\dagger$  & Conjugate transpose of an operator $A$ &  Section~\ref{sec:3}\\
        \midrule
            $\ket{\psi_c}$ & Pure state of the control qubit &  Section~\ref{sec:3}\\ 
        \midrule
            $\rho_c$ & Density operator of the control qubit  &  Section~\ref{sec:3}\\
        \midrule
            $\mathcal{S}(\cdot)(\cdot)$ & Quantum SWITCH supermap realizing a quantum path &  Section~\ref{sec:3}\\ 
		\midrule
            $\mathcal{N}_{\text{QS}}(\cdot)$ & Equivalent quantum SWITCH channel & Section~\ref{sec:3}\\
        \midrule
            $\mathcal{N}_{\text{QS}}^{\ket{\mp}}(\cdot)$ & Components of $\mathcal{N}_{\text{QS}}$ with respect to outcomes $\ket{\pm}$ of measurement of the control qubit &  Section~\ref{sec:3}\\
        \midrule
            $\mathcal{E}_p(\cdot)$ & Quantum erasure channel with erasure probability $p$ & Section~\ref{sec:4}\\
        \midrule
            $\ket{e}$ & Erasure flag of a quantum erasure channel & Section~\ref{sec:4}\\
        \midrule
            $\mathcal{P}(\cdot)$ & Quantum Pauli channel &  Section~\ref{sec:5}\\
        \midrule
            $\tilde{\mathcal{P}}$ & Probability vector of a quantum Pauli channel $\mathcal{P}$ &  Section~\ref{sec:5}\\    
        \midrule
            $H_2(X)$   & Binary Shannon entropy of the random variable $X$ & Section~\ref{sec:5}\\
        \midrule
            $Q_{\text{QS}}^{\text{UB}}$ & Upper bound on quantum capacity of the quantum path & Section~\ref{sec:5}\\
		\midrule
            $Q_{\text{QS}}^{\text{LB}}$ & Lower bound on quantum capacity of the quantum path & Section~\ref{sec:5}\\
        \midrule
            $\delta Q_{\text{QS/C}}$ & Difference between the lower bound on quantum capacity of the quantum path and the upper bound on quantum capacity of the corresponding classical path & Section~\ref{sec:5}\\
        \bottomrule
	\end{tabular}
	\caption{Adopted notation and section of the manuscript where the notation is defined or introduced.}
	\label{Tab:00}
	\hrulefill
\end{table*}

In the recent years, a lot of effort has been put to find communication scenarios allowing one to reach the rates not only beyond the classical but even the quantum Shannon's theory. Counter-intuitively, quantum mechanics allows a system to propagate simultaneously among multiple space-time trajectories \cite{ChiKri-19,KriSalEblChi-19,AbbWecHor-18}. This peculiar property enables scenarios, where quantum information carrier propagates through a quantum path, i.e., through a network path where the constituting communications channels are combined in a \textit{quantum superposition of different configurations}. The aforementioned channel placement constitutes a genuine quantum setting with no-counterpart in classical networks \cite{KouCacSim-22}.

A particular kind of quantum path is obtained through  superposition of different orders among the
constituting communication channels, as shown in Fig.~\ref{Fig:3}. In such a scenario, the causal order of the channels becomes, generally speaking, indefinite, leading to rates that can exceed the limits of the standard quantum Shannon's theory~\cite{EblSalChi-18, SalEblChi-18, ChiBanSan-18}. Therefore, it is crucial to determine the ultimate communication rates achievable by a point-to-point quantum communication protocol through noisy quantum channels, when these channels are combined in a superposition of different orders.

\subsection{Related Work}

Quantum paths, incompatible with any definite causal order of communication channels, have been experimentally realized through a quantum device called \textit{quantum SWITCH}\cite{ChiDarPer-13}. Therein, the causal order of the communication channels is controlled by a quantum degree of freedom, represented by a control qubit. Specifically, multiple implementations of the quantum SWITCH in table-top experiments have been recently proposed, e.g., photonic setups exploiting path~\cite{ProMoqAra-15, RubRozFei-17, Pro-19, GuoHuHou-18, RubTozMas-22, StrSchPet-22} or polarization~\cite{GosGiaKew-18,GosRomWhi-18,GosRom-20} of a photon as the control degree of freedom. Fig.~\ref{Fig:4} provides\footnote{We refer the reader to \cite{ChiKri-19,CalCac-19} for further details.} a schematic diagram of some of the existing photonic implementations of the quantum SWITCH.

Successful experimental implementations of the quantum SWITCH have immediately increased interest in its potential benefits. Indeed, the adoption of the quantum SWITCH has been shown to provide significant advantages for a number of problems in a wide range of fields. In particular, it leads to a computational advantage over causally ordered quantum circuits~\cite{ColDarFac-12,ChiDarPer-13,AraCosBru-14}, it allows for improved strategies for quantum metrology~\cite{ZhaYanChi-20,Liu-23} and effective discrimination of unknown quantum channels~\cite{Chi-12,BavMurQui-21}, it reduces communication complexity~\cite{FeiAraBru-15,GueFeiAra-16}, and it provides thermodynamic advantages~\cite{FelVed-20,GuhAliPar-20,SimFraGua-22,DieLisSer-23}.

While the ultimate communication rates achievable with classical paths have been deeply investigated \cite{Ben-97,PirLauOtt-17,TakGuh-14} for a large number of quantum channels, communications via quantum paths combining the communications channels in a quantum superposition of different orders caught attention of the community recently. For example, in \cite{ChaCalCac-22}, capacity of entanglement-assisted  communication over classical and quantum trajectories under quantum superdense coding protocol has been addressed. Indeed, extensions of quantum Shannon's theory to embrace quantum path and analysis of the achievable ultimate rates are currently one of the hottest research topics within the communications domain~\cite{KouCacSim-22,ChaCalCac-22, GosGiaKew-18,GuoHuHou-18,SalEblChi-18,EblSalChi-18,GosRomWhi-18,WeiNorZha-19,ChiBanSan-18,GueRubBru-19,ChiKri-19}. These efforts show that the quantum SWITCH provides advantages in communication capacity compared with classical paths for noisy channels. Moreover, the quantum SWITCH can enable noiseless communications via noisy channels, even if no information can be sent through either of the component channels individually~\cite{ChiBanSan-18,CalCac-19}. However, in the literature, analytically only a lower bound for the quantum capacity achievable through the quantum SWITCH for simple (bit- and phase-flip) channel models has been derived so far.

\subsection{Outline and Contribution}

In this paper, we address the crucial problem of estimating the  quantum capacity achievable on a quantum path, i.e., in presence of the quantum SWITCH. Against this background, our novel contributions can be summarized as follows:

\begin{itemize}
    \item for quantum erasure channels, 
     we analytically derive the quantum capacity achievable on a quantum path and we identify the region in which the quantum SWITCH incontrovertibly boosts the amount of transmissible information beyond the limits of conventional quantum Shannon's theory;
    \item for arbitrary Pauli channels, we analytically derive both a lower and an upper bound on the quantum capacity achievable on a quantum path, and we highlight three different classes of pairs of non-identical Pauli channels that, when placed in a quantum path, allow a heralded noiseless quantum transmission; also for these three classes of Pauli channels, we identify the region in which the quantum SWITCH outperforms the limits of conventional quantum Shannon's theory,
    \item for a pair of bit- and phase-flip channels, we derive a tighter lower bound than the one derived in \cite{SalEblChi-18}, and we derive, for the first time to the best of our knowledge, a closed-form expression for the upper bound depending on a computable single-letter quantity.
\end{itemize}

The theoretical analysis reveals that the adoption of the quantum SWITCH offers a substantial advantage in terms of ultimate achievable rate with respect to any classical path combining the channels in a well-defined causal order.

The rest of the paper is organized as follows. In Sec.~\ref{sec:2} we provide some preliminaries about the quantum capacity and the quantum SWITCH, and establish an upper bound on quantum capacity of the classical path. Then, in Sec.~\ref{sec:3} we derive the upper- and the lower bound on the quantum capacity in presence of the quantum SWITCH. In Sec.~\ref{sec:4}, we consider quantum erasure channels and we quantify the advantage of the quantum path over classical one. In Sec.~\ref{sec:5} we focus on Pauli channels and we quantify, in terms of achievable rates, the advantage of adopting the quantum SWITCH over classical paths. Finally in Sec.~\ref{sec:6}, we conclude the paper. In Table~\ref{Tab:00}, we summarize the notation used in this paper.


\section{Preliminaries}
\label{sec:2}

\subsection{Quantum Capacity}
\label{sec:2.1}
A fundamental property of a communication channel is its \textit{capacity}, which establishes the maximal rate of information that it can transfer under appropriately chosen encoding/decoding operations. For a quantum channel, different notions of capacity can be introduced depending on the nature of information to be transferred between the communication parties, i.e., quantum or classical~\cite{Ben-97,SmiSmoWin-08,Wil-13,Dev-05,Llo-97,SalEblChi-18}\footnote{We refer the reader to \cite{KouCacSim-22} for a concise introduction to the different notions of quantum channel capacities.}.

In the following, we restrict our attention to the transmission of quantum information, and, thus, we adopt the definition of quantum capacity $\mathcal{Q}(\mathcal{N})$ of a noisy quantum channel $\mathcal{N}$ as in \cite{SmiSmoWin-08,SalEblChi-18,Dev-05,Llo-97}. Specifically, the quantum capacity $\mathcal{Q}(\mathcal{N})$ is defined as the number of qubits transmitted per channel use in the limit of asymptotically many uses \cite{SalEblChi-18}, and it coincides with the entanglement-generating capacity as proved in \cite{Dev-05}. In order to define formally the quantum capacity $\mathcal{Q}(\mathcal{N})$, we provide the necessary notions step by step, starting with the formal definition of a \textit{quantum channel} and ending with \textit{coherent information}.

\begin{defin}[Quantum channel]
	\label{def:1}
	Let us denote with $A$ and $B$ the input and output quantum systems of a quantum communication channel $\mathcal{N}$, which is described mathematically by the completely-positive linear map \cite{BarNieSch-97}:
	\begin{equation}
		\label{eq:1}
		\mathcal{N}: \mathcal{L}(\mathbf{H}_A) \rightarrow \mathcal{L}(\mathbf{H}_B),
	\end{equation}
	with $\mathcal{L}(\mathbf{H}_A)$ and $\mathcal{L}(\mathbf{H}_B)$ are sets of density operators over the Hilbert spaces $\mathbf{H}_A$ and $\mathbf{H}_B$ of the communication parties $A$ and $B$, respectively. When channel $\mathcal{N}$ is applied to the arbitrary input density state $\rho_A \in \mathcal{L}(\mathbf{H}_A)$, the output $\mathcal{N}(\rho_A)$ is a density state $\rho_B$ belonging to $\mathcal{L}(\mathbf{H}_B)$:	
	\begin{equation}
		\label{eq:2}
		\rho_B = \mathcal{N}(\rho_A) \in \mathcal{L}(\mathbf{H}_B).
	\end{equation}
\end{defin}
It is often convenient to introduce an additional (reference) system $R$ and consider $A$ as a part of a larger system $RA$ described by an extended Hilbert space $\mathbf{H}_{RA} = \mathbf{H}_{R} \otimes \mathbf{H}_{A}$. This allows to consider $\rho_A$ as a reduced state of a pure state $\ket{\psi_{RA}} \in \mathbf{H}_{RA}$ of the larger system~\cite{SchNie-96}. This is known as purification of $\rho_A$ and leads to the following formal definition.

\begin{defin}[Purification]
	\label{def:2}
	A pure state $\ket{\psi_{RA}}$ is a \textit{purification} of a state $\rho_A$ if:
		\begin{equation}
		\label{eq:3}
		\rho_A = \text{Tr}_R \Bigl[\ket{\psi_{RA}}\bra{\psi_{RA}}\Bigr],
	\end{equation}
	for a certain reference system $R$, with $\text{Tr}_R[\cdot]$ denoting the partial trace operator with respect to $\mathbf{H}_{R}$.
\end{defin}

In terms of channel $\mathcal{N}$ effects -- by introducing the purification of the state $\rho_A$ -- we can state that the joint system $RA$ evolves according to the ``extended'' super-operator $(\mathcal{I}_R \otimes \mathcal{N})$, where $\mathcal{I}_R$ is an identity channel for the reference system $R$ \cite{SchNie-96}, by producing the output state $\rho_{RB}$:
\begin{equation}
	\label{eq:4}
	\rho_{RB}= (\mathcal{I}_R \otimes \mathcal{N}) (\rho_{RA}).
\end{equation}

\begin{defin}[Choi matrix]
	\label{def:3}
	When the Hilbert space $\mathbf{H}_R$ of the reference system $R$ has the same dimension $d$ as $\mathbf{H}_A$, and the state $\ket{\psi_{RA}}$ of the joint system $RA$ is a bipartite maximally entangled\footnote{We observe that usually the Choi matrix is defined using an unnormalized entangled state $\sum_i \ket{i}\otimes\ket{i}$. In this paper we adopted a normalized entangled state $\frac{1}{\sqrt{d}}\sum_i \ket{i}\otimes\ket{i}$ for the sake of simplicity, so that the corresponding Choi matrix can be regarded as a normalized quantum state.} -- i.e., $\ket{\psi_{RA}} = \ket{\Phi^+} \eqdef \frac{1}{\sqrt{d}} \sum_i \ket{i} \otimes \ket{i}$ -- the output state $\rho_{RB}$ is known as the \textit{Choi matrix} $\rho_{\mathcal{N}}$ of the channel $\mathcal{N}$ \cite{NieChu-11}:
	\begin{equation}
		\label{eq:5}
		\rho_{\mathcal{N}} \eqdef \rho_{RB}\mid_{\rho_{RA}=\rho_{\Phi^+}}=(\mathcal{I}_R \otimes \mathcal{N}) (\rho_{\Phi^+}),
	\end{equation}
	with $\rho_{\Phi^+} \eqdef \ket{\Phi^+}\bra{\Phi^+}$.
\end{defin}

\begin{defin}[Choi-stretchable channel]
	\label{def:3.5}
	A quantum channel $\mathcal{N}$ is called Choi-stretchable if its action can be reproduced by local operatios and classical communications (LOCC) $\mathcal{T}(\cdot)$ over its Choi matrix $\rho_\mathcal{N}$~\cite{PirLauOtt-17}, i.e.:
	\begin{equation}
	    \mathcal{N}(\rho) = \mathcal{T}(\rho \otimes \rho_\mathcal{N}).
	\end{equation}
    Paradigmatic examples of Choi-stretchable channels are erasure and Pauli channels~\cite{PirLauOtt-17}, which are analyzed in Section~\ref{sec:4} and Section~\ref{sec:5}, respectively.
\end{defin}

\begin{defin}[Coherent Information]
	\label{def:4}
	The coherent information\footnote{The coherent information is often also denoted with the symbol $I_c(A\rangle B)_{\rho_{RB}}$ \cite{Wil-13}.} $I_c(\rho,\mathcal{N})$ of the channel $\mathcal{N}$ with respect to the arbitrary input state $\rho$ is defined as \cite{SchNie-96,Llo-97,Dev-05}:
	\begin{equation}
		\label{eq:6}
		I_c(\rho,\mathcal{N}) = S(\rho_B)-S(\rho_{RB}),
	\end{equation}
	where $\rho_B$ and $\rho_{RB}$ are defined in \eqref{eq:2} and \eqref{eq:4}, respectively, and $S(\sigma) \eqdef - \text{Tr}[\sigma \log_2{\sigma}]$ denotes the \textit{von Neumann entropy} of the considered system state $\sigma$. 
\end{defin}

Intuitively, the coherent information $I_c(\rho,\mathcal{N})$ aims at describing -- by subtracting the von Neumann \textit{entropy exchange} $S(\rho_{RB})$ between the input state and the channel from the von Neumann entropy of the output state $S(\rho_B)$ -- the amount of quantum information preserved after the state $\rho$ goes through the channel $\mathcal{N}$. Stemming from \eqref{eq:6}, we can now define the coherent information of the channel $\mathcal{N}$ \cite{SchNie-96,Llo-97,Dev-05,DevWin-05,SalEblChi-18}.

\begin{defin}[Channel Coherent Information]
	\label{def:5}
	The coherent information $I_c(\mathcal{N})$ of the channel $\mathcal{N}$ is defined as:
	\begin{align}
		\label{eq:7}
		I_c(\mathcal{N})=\max_{\ket{\psi}_{RA}} I_c(\rho,\mathcal{N}),
	\end{align}
	where the maximum is taken with respect to all pure bipartite states $\ket{\psi}_{RA}$ that are purification of the input state $\rho$.
\end{defin}

The channel coherent information -- often referred to as one-shot capacity -- plays a crucial role in the definition of the quantum capacity \cite{SchNie-96,Llo-97,Dev-05}, analogous to the role played by the (classical) mutual information in classical information theory. However, the coherent information exhibits some nasty properties. In fact, the coherent information can be negative and, in general, it is not additive. Hence, differently from classical Shannon's theory, the quantum capacity cannot be determined by simply evaluating the one-shot capacity\footnote{Exception is constituted by the class of degradable channels, whose quantum capacity is exactly equal to the coherent information, i.e., to the one-shot capacity. We refer the reader to \cite{GyoImrNgu-18} for a technical and historical survey about quantum capacity formulation.}, but it generally requires the asymptotic formulation given below, which holds in the region where the coherent information is non-negative \cite{SchNie-96,Llo-97,Dev-05,DevWin-05}.

\begin{defin}[Quantum Channel Capacity]
	\label{def:6}
	The quantum channel capacity is given by:
	\begin{equation}
		\label{eq:8}
		\mathcal{Q}(\mathcal{N})= \lim_{n \rightarrow +\infty} \frac{1}{n} I_c(\mathcal{N}^{\otimes n}),
	\end{equation}
	where $I_c(\mathcal{N})$ is defined in \eqref{eq:7} and $\mathcal{N}^{\otimes n}$ denotes $n$ uses of channel $\mathcal{N}$.
\end{defin}

From Definition~\ref{def:6}, it is evident that the evaluation of the quantum capacity is not a trivial task. In fact, the expression \eqref{eq:8} requires to maximize the coherent information over an unbounded number of channel uses. 
Nevertheless, in general, $I_c(\mathcal{N})$ constitutes a lower bound for the quantum capacity, and we will exploit this property in Sec.~\ref{sec:3} to derive the capacity region of a quantum path implemented via the quantum SWITCH.

Before to proceed further, we provide two more definitions that we will use to estimate the quantum capacity of a communication channel.
\begin{defin}[Quantum Relative Entropy]\label{def:QRE}
Quantum relative entropy of a state $\rho$ with respect to the state $\sigma$ is defined as \cite{Wil-13}:
\begin{align}
	\label{def:6A}
	S(\rho||\sigma) &\eqdef \text{Tr}\left[\rho\left(\log_2\rho - \log_2 \sigma\right)\right]. 
\end{align}
\end{defin}

\begin{defin}[Relative Entropy of Entanglement]\label{def:REE}
Relative entropy of entanglement of a state $\rho$ is defined as its minimal quantum relative entropy with respect to a separable state \cite{Wil-13}:
\begin{equation}
	\label{def:6B}
	E_R(\rho) \eqdef \inf_{\sigma_s} S(\rho||\sigma_s),
\end{equation}
where $\sigma_s$ is an arbitrary separable state.
\end{defin}

\subsection{Quantum Capacity through Classical paths}
\label{sec:2.2}

In order to highlight the benefits of exploiting quantum paths for communicating, we question firstly the best performance of quantum noisy channels placed into a classical path. Specifically, we consider the following communication model: a quantum message to be transmitted at the destination through a channel $\mathcal{N}_{\text{C}}$ that is a cascade of noisy quantum channels (for example, corresponding to wires of different quality). For the sake of simplicity, we assume that the cascade consists of two quantum channels, denoted with $\mathcal{D}$ and $\mathcal{E}$, respectively. In such a cascade, as shown in Fig.~\ref{Fig:1}, two possible alternative configurations can be realized:
\begin{itemize}
	\item[-] channel $\mathcal{E}$ is traversed after channel $\mathcal{D}$, giving rise to the classical path $\mathcal{D} \rightarrow \mathcal{E}$;
	\item[-] or vice versa, giving rise to the classical path $\mathcal{E} \rightarrow \mathcal{D}$.
\end{itemize}
Since $\mathcal{N}_{\text{C}}(\mathcal{D}, \mathcal{E})$ establishes a classical path, it can either a-priori realize one of the mentioned configurations or choose between them randomly. Either way, when the two channels are traversed in a well-defined order, the bottleneck inequality holds. This means that the overall quantum capacity $\mathcal{Q}(\mathcal{N}_{\text{C}})$ associated with the considered classical path is smaller than the minimum between the individual capacities $\mathcal{Q}(\mathcal{D})$ and $ \mathcal{Q}(\mathcal{E})$ \cite{Wil-13,SalEblChi-18,CalCac-19}. This establishes the upper bound on the quantum capacity achievable on a classical path:
	\begin{equation}
		\label{eq:9}
		\mathcal{Q}(\mathcal{N}_{\text{C}}) \leq \min\{ \mathcal{Q}(\mathcal{D}), \mathcal{Q}(\mathcal{E}) \} .
	\end{equation}

\section{Quantum paths via Quantum Switch}
\label{sec:3}
Let us assume, without loss of generality, the message being a qubit $\ket{\varphi} \in \mathbf{H}_A$ (where $\mathbf{H}_A$ denotes the associated Hilbert space) with density matrix $\rho \eqdef \ket{\varphi}\bra{\varphi}$. As mentioned in Section~\ref{sec:1}, the quantum SWITCH is a device that implements a quantum path by allowing a quantum information carrier -- a qubit $\ket{\varphi}$ -- to experience a set of evolutions in a superposition of alternative orders \cite{SalEblChi-18,EblSalChi-18,CalCac-19}. In the quantum SWITCH, the causal order between the channels is determined by a quantum degree of freedom, represented by the \textit{control qubit} $\ket{\varphi_c}$.

Specifically, if the control qubit is initialized to the basis state $\ket{\varphi_c} = \ket{0}$, the quantum SWITCH enables the message $\ket{\varphi}$ to propagate through the classical path $\mathcal{D \rightarrow E}$, representing channel $\mathcal{E}$ being traversed after channel $\mathcal{D}$. Similarly, if the control qubit is initialized to the other basis state $\ket{\varphi_c} = \ket{1}$, the quantum SWITCH enables the message $\ket{\varphi}$ to propagate through the alternative classical path $\mathcal{E \rightarrow D}$, representing channel $\mathcal{E}$ being traversed before channel $\mathcal{D}$.

Differently, if the control qubit is initialized to a state different from $\ket{0}$ or $\ket{1}$ (hence, to a certain superposition of them), the message $\ket{\varphi}$ propagates through a quantum path. For example, for $\ket{\varphi_c} = \ket{\pm} \equiv \frac{\ket{0} \pm \ket{1}}{\sqrt{2}}$, the message experiences an equally balanced superposition of the two alternative evolutions $\mathcal{D \rightarrow E}$ and $\mathcal{E \rightarrow D}$ as shown in Fig.~\ref{Fig:3}. The channel placement enabled by the quantum SWITCH is genuinely quantum and, as shown in the following, provides a non-trivial resource for the channel capacity activation that cannot be reproduced through a classical channel placement\footnote{It important to highlight that, in the described classical and quantum placements of channels, both channels $\mathcal{E}$ and $\mathcal{D}$ are used in a certain combination. This should not be confused with settings utilized for classical capacity activation where only one of the channels is used at time and a non-zero capacity of the equivalent channel is achieved by encoding the information in whether one or the other channel has been used~\cite{CovTho-06}.}.   

\begin{figure}[t]
	\centering
	\includegraphics[width=\columnwidth]{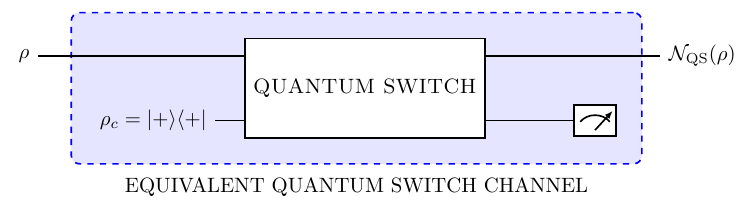}
	\caption{Equivalent channel model of a quantum path implemented by utilizing the quantum SWITCH, mapping the input density matrix $\rho$ into the output density matrix $\mathcal{N}_{\text{QS}}(\rho)$. The control qubit $\ket{\varphi_c}$ is initialized to $\ket{+}$, and the expression of $\mathcal{N}_{\text{QS}}(\rho)$ depends on the measurement outcome of the control qubit at the output of the quantum SWITCH.} \label{Fig:6}
	\hrulefill
\end{figure}

The behavior of the quantum SWITCH can be mathematically described by a higher-order operation (supermap) that sends quantum channels to a quantum channel that is not necessarily compatible with any well-defined order of their occurrence~\cite{SalEblChi-18,ChiKri-19,CalCac-19}. In particular, for two input quantum channels $\mathcal{D}$ and $\mathcal{E}$, the output quantum channel is defined as:
\begin{equation}
	\label{eq:15}
	\mathcal{S}(\mathcal{D}, \mathcal{E},\rho_c)(\rho) = \sum_{i,j} W_{ij} (\rho \otimes \rho_c) W_{ij}^\dag,
\end{equation}
where $\rho \eqdef \ket{\varphi}\bra{\varphi}$ and $\rho_c \eqdef \ket{\varphi_c}\bra{\varphi_c}$, and where $\{ W_{ij} \}$ denotes the set of Kraus operators associated with the quantum path, given by \cite{SalEblChi-18,ChiKri-19}:
\begin{equation}
	\label{eq:16}
	W_{ij} = E_j D_i \otimes \ket{0}\bra{0}_c + D_i E_j \otimes \ket{1}\bra{1}_c.
\end{equation}
In \eqref{eq:16}, $\{D_i\}$ and $\{E_j\}$ are the Kraus operators of the channels $\mathcal{D}$ and $\mathcal{E}$, respectively.

When the control qubit is initialized in the state $\ket{\varphi_c} = \ket{+}$, the quantum SWITCH maps the channels $\mathcal{D}$ and $\mathcal{E}$ to the channel given by \cite{ChiBanSan-18}:
\begin{align}
    \label{Eq:15}
    \nonumber \mathcal{S}&(\mathcal{D}, \mathcal{E}, \ket{+}\bra{+})(\rho) = \\
    \nonumber &= p_{-} \,\mathcal{N}_{\text{QS}}^{\ket{-}}(\rho) \otimes \ket{-}\bra{-} + p_{+}\,\mathcal{N}_{\text{QS}}^{\ket{+}}(\rho) \otimes \ket{+}\bra{+}  \\
    &+ \chi(\rho) \otimes \ket{-}\bra{+} + \chi^\dagger(\rho) \otimes \ket{+}\bra{-},
\end{align}
where
\begin{eqnarray}
    \label{Eq:16}
    \mathcal{N}_{\text{QS}}^{\ket{\mp}}(\rho) = \frac{1}{4p_\mp} \sum_{i,j} [ D_i, E_j ]_{\mp} \rho [ D_i, E_j]_{\mp}^\dagger, \\
    \label{Eq:17}
    p_{\pm} = \frac{1}{4} \operatorname{Tr}\Bigl[\sum_{i,j} [ D_i, E_j ]_{\mp} \rho [ D_i, E_j]_{\mp}^\dagger\Bigr],
    \end{eqnarray}
with $[D_i, E_j]_\mp \eqdef D_i E_j \mp E_j D_j$ denoting the commutator ($-$) or anti-commutator ($+$) of\footnote{We observe that usually the anti-commutator of $A$ and $B$ is denoted with the symbol $\{A,B\}$. In this paper we adopted a different notation for the sake of conciseness, by avoiding to specialize equations~\eqref{Eq:16} and \eqref{Eq:17} twice.} $D_i$ and $E_j$, and 
\begin{eqnarray}
    \chi(\rho) = \frac{1}{4} [ D_i, E_j ]_- \rho [ D_i, E_j]_+^\dagger.
\end{eqnarray}

Performing a measurement of the control qubit in the Hadamard basis, one obtains a mixture of the states $\mathcal{N}_{\text{QS}}^{\ket{\mp}}(\rho)$ with the corresponding probabilities $p_\mp$. In this way, it is convenient to introduce the \textit{equivalent channel} model of a quantum path implemented by utilizing the quantum SWITCH, as depicted in Fig.~\ref{Fig:6}. The input and the output of this equivalent channel are, respectively, the original quantum state $\rho$ and $\mathcal{N}_{\text{QS}}(\rho)$, denoting\footnote{We omit the dependence of $\mathcal{N}_{\text{QS}}(\cdot)$ on $\mathcal{D}$ and $\mathcal{E}$ for the sake of notation simplicity.} the quantum state after the measurement process on the control qubit at the output of the quantum SWITCH.

Indeed, since the measurement process is characterized by two distinct outcomes, $\ket{-}$ and $\ket{+}$ occurring with probability $p_-$ and $p_+$, and in light of the aforementioned equivalent channel model, we can associate $\mathcal{N}_{\text{QS}}^{\ket{+}}(\rho)$ with the channel output when the measurement process on the control qubit $\ket{\varphi_c}$ returns $\ket{+}$. Conversely, we associate $\mathcal{N}_{\text{QS}}^{\ket{-}}(\rho)$ with the channel output when the measurement process on the control qubit $\ket{\varphi_c}$ returns $\ket{-}$. By accounting for this equivalent model, it is possible to write the input-output relationship for the equivalent channel $\mathcal{N}_{\text{QS}}(\cdot)$ as:
\begin{equation}
	\label{Eq:19}
		\mathcal{N}_{\text{QS}}(\rho) = \begin{cases}
				\mathcal{N}_{\text{QS}}^{\ket{-}}(\rho) & \text{with prob. } p_-,\\
				\displaystyle \mathcal{N}_{\text{QS}}^{\ket{+}}(\rho) & \text{with prob. } p_+.
		\end{cases}
\end{equation}

Therefore, the quantum capacity of the equivalent quantum SWITCH  is an average quantum capacity over the channels heralded with measurement outcomes $\ket{-}$ and $\ket{+}$, respectively:
\begin{equation}
    \label{Eq:20}
    \mathcal{Q}(\mathcal{N}_{\text{QS}}) = p_- \mathcal{Q}(\mathcal{N}_{\text{QS}}^{\ket{-}}) + p_+ \mathcal{Q}(\mathcal{N}_{\text{QS}}^{\ket{+}}).
\end{equation}
This allows one to estimate the lower bound of quantum capacity of a quantum path implemented via the quantum SWITCH when it cannot be calculated explicitly, as stated with the following proposition.

\begin{prop}
	\label{prop:1}
	The quantum capacity $\mathcal{Q}(\mathcal{N}_{\text{\rm QS}})$ of the equivalent quantum SWITCH channel $\mathcal{N}_{\text{\rm QS}}(\cdot)$ is lower-bounded as follows:
	\begin{eqnarray}
	    \label{Eq:22}
	   \nonumber \mathcal{Q}(\mathcal{N}_{\text{\rm QS}}) &\geq& \max\{0, p_- I_c(\mathcal{N}_{\text{\rm QS}}^{\ket{-}}) \} \\ 
	    &+&  \max\{0, p_+ I_c(\mathcal{N}_{\text{\rm QS}}^{\ket{+}}) \}.
	\end{eqnarray}	
	\begin{IEEEproof}
		See Appendix~\ref{App:1}
	\end{IEEEproof}
\end{prop}

On the other hand, the equivalent quantum SWITCH channel~\eqref{Eq:19} can be represented by its Choi matrix $\rho_{\mathcal{N}_{\text{QS}}}$ that is given by:
\begin{equation}
    \label{eq:ChoiSwitch}
	\rho_{\mathcal{N}_{\text{QS}}} =
		\begin{cases}
			\rho_{\mathcal{N}_{\text{QS}}^{\ket{-}}} & \text{with prob. } p_-,\\
			\rho_{\mathcal{N}_{\text{QS}}^{\ket{+}}} & \text{with prob. } p_+,
		\end{cases}
\end{equation}
where $\rho_{\mathcal{N}_{\text{QS}}^{\ket{\mp}}}$ denotes the Choi matrix of the components $\mathcal{N}_{\text{QS}}^{\ket{\mp}}(\cdot)$ of the equivalent quantum SWITCH channel \eqref{Eq:19} when the measurement outcomes are $\ket{-}$ and $\ket{+}$, respectively. If $\mathcal{N}_{\text{QS}}^{\ket{\mp}}(\cdot)$ are Choi-stretchable, their representation through Choi matrices $\rho_{\mathcal{N}_{\text{QS}}^{\ket{\mp}}}$ allows one to estimate the upper bound of quantum capacity of a quantum path implemented via the quantum SWITCH, as stated with the following proposition.

\begin{prop}
	\label{prop:2}
	If both $\mathcal{N}_{\text{\rm QS}}^{\ket{\mp}}$ are Choi-stretchable, the quantum capacity $\mathcal{Q}(\mathcal{N}_{\text{\rm QS}})$ of the equivalent quantum SWITCH channel $\mathcal{N}_{\text{\rm QS}}(\cdot)$ is upper-bounded as follows:
	\begin{equation}
	    \label{Eq:23}
        \mathcal{Q}(\mathcal{N}_{\text{\rm QS}}) \leq p_- E_R(\rho_{\mathcal{N}_{\text{\rm QS}}^{\ket{-}}}) + p_+ E_R(\rho_{\mathcal{N}_{\text{\rm QS}}^{\ket{+}}}).
    \end{equation}
    \begin{IEEEproof}
		See Appendix~\ref{App:2}
	\end{IEEEproof}
\end{prop}

\begin{figure*}[t]
    \begin{minipage}[t] {0.49\textwidth}
        \centering
        \includegraphics[width=\columnwidth]{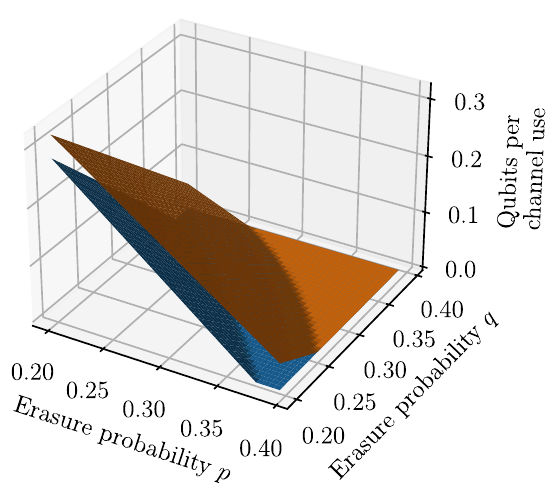}
	    \caption{Three-dimensional plot of the quantum capacity of classical (blue surface) and quantum (orange surface) paths as a function of the error probabilities $p$ and $q$ of the quantum erasure channels $\mathcal{D} = \mathcal{E}_p$ and $\mathcal{E} = \mathcal{E}_q$.}
	    \label{Fig:A5.1}
    \end{minipage}
    \hspace{0.02\textwidth}
    \begin{minipage}[t]{0.49\textwidth}
        \centering
        \includegraphics[width=\columnwidth]{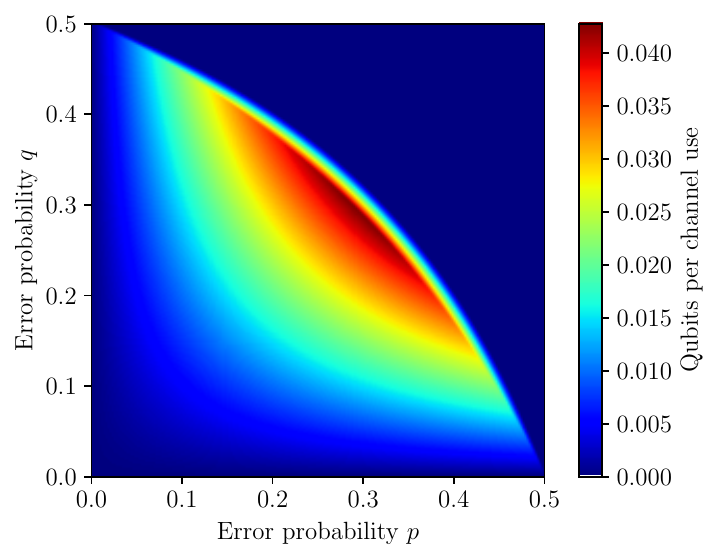}
	    \caption{Density plot of the difference between the quantum SWITCH capacity and the classical path capacity as a function of the error probabilities $p$ and $q$ of the quantum erasure channels $\mathcal{D} = \mathcal{E}_p$ and $\mathcal{E} = \mathcal{E}_q$. The rationale for the null difference in the upper right part of the graph becomes evident by noting in Fig.~\ref{Fig:A5.1} that both the capacities fall to zero in the considered region.}
	\label{Fig:A5.2}
    \end{minipage}
    \label{Fig:A5}
    \hrulefill
\end{figure*}

\section{Erasure channels}
\label{sec:4}

We start by considering quantum erasure channels, which represent one of the simplest yet useful models of noisy quantum communication. On the one hand, it is one of a few communication channels whose quantum capacity can be computed exactly~\cite{Ben-97}. On the other hand, it captures erasure errors which cannot be modelled by Pauli channels considered in Section~\ref{sec:5}, namely, loss of information carrier or information leakage to other states~\cite{MurZouLi-17}. At a practical level, quantum erasure channel model is widely adopted  in quantum satellite communications~\cite{SidJosGun-21,Zho-23}.

Technically, a quantum erasure channel $\mathcal{E}_p$ either transmits the input qubit $\rho \in \mathcal{L}(\mathbf{H}_A)$ faithfully or it replaces it with a probability $p$ by the erasure flag $\ket{e}$, that is orthogonal to $\mathbf{H}_A$:
\begin{equation}
    \mathcal{E}_p(\rho) = (1-p)\rho + p\ket{e}\bra{e}.
\end{equation}
The quantum capacity of $\mathcal{E}_p$ is non-zero for any $0 \leq p < \frac{1}{2}$, and it is given by~\cite{Ben-97}:
\begin{equation}
    \mathcal{Q}(\mathcal{E}_p) = \max\{0, 1-2p\}.
\end{equation}

\subsection{Classical path}
\label{sec:4.1}

Now we focus on a cascade of two quantum erasure channels $\mathcal{D} = \mathcal{E}_p$ and $\mathcal{E} = \mathcal{E}_q$ characterized by not necessarily equal erasure probabilities $p$ and $q$, respectively. Any classical path through these two channels establishes the same overall channel given by:
\begin{equation}
    \label{Eq:26}
    \mathcal{N}_{\text{C}}(\mathcal{E}_p, \mathcal{E}_q)(\rho) = (1-p)(1-q) \rho + (p + q - pq) \ket{e}\bra{e},
\end{equation}
which, in turn, denotes a quantum erasure channel with (increased) erasure probability $p+q-pq$. Therefore, its quantum capacity is given by:
\begin{equation}
    \label{Eq:27}
    \mathcal{Q}(\mathcal{N}_{\text{C}}) = \max\{0, 1-2(p+q-pq)\}.
\end{equation}

\subsection{Quantum path}
\label{sec:4.2}
When the application of quantum erasure channels is controlled via the quantum SWITCH, the following equivalent channel arises:
\begin{align}
    \label{Eq:28}
    \nonumber \mathcal{S}&(\mathcal{E}_p, \mathcal{E}_q, \ket{+}\langle+|)(\rho) =\\ 
    \nonumber & \Biggl( (1-p)(1-q)\rho + \Bigl(p + q - \frac{3pq}{2}\Bigr) \ket{e}\bra{e} \Biggr) \otimes  \ket{+}\langle+| \\
    &+ \frac{pq}{2}\ket{e}\bra{e} \otimes \ket{-}\langle-|.
\end{align}
Therefore, the components of the quantum SWITCH equivalent channel $\mathcal{N}_{\text{QS}}(\cdot)$ read:
\begin{eqnarray}
    \mathcal{N}_{\text{QS}}^{\ket{-}}(\rho) &=& \ket{e}\bra{e}, \\
    \nonumber \mathcal{N}_{\text{QS}}^{\ket{+}}(\rho) &=& \frac{1}{p_+} \Bigl( (1-p)(1-q)\rho \\
    &+& \Bigl(p + q - \frac{3pq}{2}\Bigr) \ket{e}\bra{e} \Bigr),
\end{eqnarray}
with probabilities $p_- = \frac{pq}{2}$ and $p_+ = 1 - \frac{pq}{2}$, respectively. It is easy to recognize that both components of the equivalent channel are in turn quantum erasure channels.

Interestingly, the channel associated with the outcome $\ket{-}$ of the control qubit measurement has the erasure probability $1$, i.e., transmits no information. On the other hand, the channel associated with the outcome $\ket{+}$ exhibits an erasure probability smaller than one of the classical path $\mathcal{N}_{\text{C}}$. In this way, in the quantum path, a part of the erasure probability of the cascade of erasure channels is ``separated''.

Since the component $\mathcal{N}_{\text{QS}}^{\ket{-}}$ transmits no information, the quantum capacity of the equivalent channel`$\mathcal{N}_{\text{QS}}(\cdot)$ is completely determined by the component $\mathcal{N}_{\text{QS}}^{\ket{+}}$:
\begin{eqnarray}
    \label{Eq:29}
    \nonumber \mathcal{Q} (\mathcal{N}_{\text{QS}}) &=& p_+ \mathcal{Q} (\mathcal{N}_{\text{QS}}^{\ket{+}}) \\
    &=& \max\Bigl\{0, 1-2\Bigl(p+q-\frac{3pq}{2}\Bigr)\Bigr\},
\end{eqnarray}
which extends the known result on capacity of quantum path for erasure channels with $p=q$~\cite{ChiKri-19}. Quantum capacities of classical and quantum paths \eqref{Eq:27} and \eqref{Eq:29} are plotted on Fig.~\ref{Fig:A5.1}, whereas the difference $\mathcal{Q} (\mathcal{N}_{\text{QS}}) - \mathcal{Q} (\mathcal{N}_{\text{C}})$ is plotted on Fig.~\ref{Fig:A5.2} and its maximum $\frac{1}{4}(3 - 2\sqrt{2})$ can be achieved for two identical quantum erasure channels with $p = q = 1 - \frac{\sqrt{2}}{2}$.

\begin{rem}
    It is worthwhile to note that there exist quantum erasure channels for which a quantum path exhibits non-zero quantum capacity, whereas any classical path exhibits null capacity. This means that a quantum path activates the transmission of quantum information through quantum erasure channels, whose any classical configuration does not transmit it at all. The erasure probabilities of such pairs of erasure channels must satisfy the following condition:
    \begin{equation}
        \label{Eq:30}
        0 < \frac{1-2q}{2(1-q)} < p < \frac{1-2q}{2-3q} < \frac{1}{2}.
    \end{equation}
\end{rem}

\section{Pauli channels}
\label{sec:5}
Pauli channels constitute paradigmatic models in quantum communications, since they are able to capture bit-flip errors, as well as uniquely quantum features of noise such as phase-flip errors and its combination with bit-flip errors. These are the basic errors that can occur in practice~\cite{PirLauOtt-17} in noisy transmissions of qubits.
Indeed, the ubiquitous presence of quantum noise that can be modelled with Pauli channels has been validated in a number of works, e.g.~\cite{PirLauOtt-17,CacCalVanHan-19,ChaCalCac-22}, including Google's one about quantum supremacy \cite{Arute-19}. Therein, the action of noise is modelled as follows:
\begin{equation}
	\label{Eq:31}
	\mathcal{P}(\rho) = p_0 \rho + p_X X \rho X + p_Y Y \rho Y + p_Z Z \rho Z,
\end{equation}
with $X = \begin{bmatrix}0 & 1\\1 & 0\end{bmatrix}$, $Y = \begin{bmatrix}0 & -i\\i & 0\end{bmatrix}$, and $Z = \begin{bmatrix}1 & 0\\0 & -1\end{bmatrix}$ denoting the Pauli gates \cite{NieChu-11}.

Overall, a Pauli channel $\mathcal{P}$ is characterized by the corresponding probability vector $(p_0, p_X, p_Y, p_Z)$, the components whereof sum up to unity. Therefore, for the sake of simplicity, we characterize a Pauli channel $\mathcal{P}$ by its probability vector $\tilde{\mathcal{P}}$,
\begin{equation}
    \label{Eq:32}
    \tilde{\mathcal{P}} = (p_0, p_X, p_Y, p_Z).
\end{equation}

\subsection{Arbitrary Pauli channels}
\label{sec:5.1}

We consider two not necessarily identical Pauli channels $\mathcal{D}$ and $\mathcal{E}$ with probability vectors $\tilde{\mathcal{D}} = (p_0^{\mathcal{D}}, p_X^{\mathcal{D}}, p_Y^{\mathcal{D}}, p_Z^{\mathcal{D}})$ and $\tilde{\mathcal{E}} = (p_0^{\mathcal{E}}, p_X^{\mathcal{E}}, p_Y^{\mathcal{E}}, p_Z^{\mathcal{E}})$, respectively. A quantum path leads to the equivalent quantum SWITCH channel characterized by the following probability vectors:
\begin{eqnarray}
    \label{eq:PauliGenCompMinus}
    \tilde{\mathcal{N}}_{\text{QS}}^{\ket{-}} &=& \frac{1}{p_-}\Bigl(0, \eta_{YZ}, \eta_{XZ}, \eta_{XY} \Bigr), \\
    \label{eq:PauliGenCompPlus}
    \tilde{\mathcal{N}}_{\text{QS}}^{\ket{+}} &=& \frac{1}{p_+}\Bigl(\sum_\alpha \eta_{\alpha \alpha}, \eta_{0X}, \eta_{0Y}, \eta_{0Z} \Bigr),
\end{eqnarray}
where $\alpha \in \{0,X,Y,Z\}$ are defined by summing the components of the corresponding probability vectors, and $\eta_{\alpha\beta} = p_\alpha^{\mathcal{D}} p_\beta^{\mathcal{E}} + p_\beta^{\mathcal{D}} p_\alpha^{\mathcal{E}}$. The measurement outcome probabilities read $p_- = \eta_{YZ} + \eta_{XZ} + \eta_{XY}$ and $p_+ = \sum_\alpha \eta_{\alpha \alpha} + \eta_{0X} + \eta_{0Y} + \eta_{0Z}$.

Stemming from  Propositions~\ref{prop:1} and \ref{prop:2}, and denoting $\eta = \eta_{XY} + \eta_{YZ} + \eta_{XZ}$, $\tilde{p}_\alpha = \tilde{p}_\alpha^{\mathcal{E}} + \tilde{p}_\alpha^{\mathcal{D}} - 2\tilde{p}_\alpha^{\mathcal{E}}\tilde{p}_\alpha^{\mathcal{D}}$ for any $\alpha \in \{X,Y,Z\}$, and $\tilde{p} = \tilde{p}_X + \tilde{p}_Y + \tilde{p}_Z$, we obtain the following lower and upper bounds of the quantum capacity achievable with a quantum path.

\begin{prop}
	\label{prop:3}
	The quantum capacity $\mathcal{Q}(\mathcal{N}_{\text{\rm QS}})$ of the equivalent quantum SWITCH channel $\mathcal{N}_{\text{\rm QS}}(\cdot)$ for arbitrary Pauli channels is lower-bounded by $\mathcal{Q}_{\text{\rm QS}}^{\text{\rm LB}}$ as follows:
	\begin{eqnarray}
	\label{Eq:37}
    \nonumber \mathcal{Q}(\mathcal{N}_{\text{\rm QS}}) &\geq& \max\Bigl\{0, \eta(1-\log_2\eta) + \eta_{XY} \log_2\eta_{XY} \\
    \nonumber  &+& \eta_{YZ} \log_2\eta_{YZ} + \eta_{XZ} \log_2\eta_{XZ}\Bigr\} \\
    \nonumber  &+& \max\Bigr\{ 0, (1-\eta)(1-\log_2(1-\eta)) \\
    \nonumber  &+& (1-\tilde{p}+\eta) \log_2(1-\tilde{p}+\eta) \\
    \nonumber  &+& (\tilde{p}_X - \eta_{XY} - \eta_{XZ})\log_2(\tilde{p}_X - \eta_{XY} - \eta_{XZ}) \\
    \nonumber &+& (\tilde{p}_Y - \eta_{XY} - \eta_{YZ})\log_2(\tilde{p}_Y - \eta_{XY} - \eta_{YZ}) \\
    \nonumber &+& (\tilde{p}_Z - \eta_{XZ} - \eta_{YZ})\log_2(\tilde{p}_Z - \eta_{XZ} - \eta_{YZ}) \Bigr\} \\
    &\eqdef& \mathcal{Q}_{\text{\rm QS}}^{\text{\rm LB}},
\end{eqnarray}
	\begin{IEEEproof}
		See Appendix~\ref{App:3}.
	\end{IEEEproof}
\end{prop}

\begin{prop}
	\label{prop:4}
	The quantum capacity $\mathcal{Q}(\mathcal{N}_{\text{\rm QS}})$ of the equivalent quantum SWITCH channel $\mathcal{N}_{\text{\rm QS}}(\cdot)$ for arbitrary Pauli channels is upper-bounded by $\mathcal{Q}_{\text{\rm QS}}^{\text{\rm UB}}$ as follows:
\begin{eqnarray}
    \label{Eq:38}
    \nonumber \mathcal{Q}(\mathcal{N}_{\text{\rm QS}}) &\leq& 1 + \eta_{YZ} \log_2\eta_{YZ} + \eta_{XZ} \log_2\eta_{XZ} \\
    \nonumber  &-& (\eta_{YZ} + \eta_{XZ}) \log_2(\eta_{YZ} + \eta_{XZ}) \\
    \nonumber  &+& (1 - \tilde{p} + \eta) \log_2(1 - \tilde{p} + \eta) \\
    \nonumber  &-& (1 - \tilde{p}_X - \tilde{p}_Y + \eta_{XY}) \\
    \nonumber &\cdot& \log_2(1 - \tilde{p}_X - \tilde{p}_Y + \eta_{XY}) \\
    \nonumber  &-& (\tilde{p}_X + \tilde{p}_Y - \eta_{XY} - \eta) \\
    \nonumber &\cdot& \log_2(\tilde{p}_X + \tilde{p}_Y - \eta_{XY} - \eta) \\
    \nonumber  &+& (\tilde{p}_X - \eta_{XY} - \eta_{XZ}) \log_2(\tilde{p}_X - \eta_{XY} - \eta_{XZ}) \\
    \nonumber &+& (\tilde{p}_Y - \eta_{XY} - \eta_{YZ}) \log_2(\tilde{p}_Y - \eta_{XY} - \eta_{YZ})\\
    \nonumber &+& (\tilde{p}_Z - \eta_{XZ} - \eta_{YZ}) \log_2(\tilde{p}_Z - \eta_{XZ} - \eta_{YZ}) \\
    &\eqdef& \mathcal{Q}_{\text{\rm QS}}^{\text{\rm UB}}.
\end{eqnarray}
	\begin{IEEEproof}
		See Appendix~\ref{App:4}.
	\end{IEEEproof}
\end{prop}

Since Pauli channels are commutative, any classical path through these two channels has the same output, which can be characterized through the probability vector given by:
\begin{equation}
    \label{eq:PauliGenClass}
    \tilde{\mathcal{N}}_{\text{C}} = \Bigl(\sum_{\alpha} \eta_{\alpha \alpha}, \eta_{0X} + \eta_{YZ}, \eta_{0Y} + \eta_{XZ}, \eta_{0Z} + \eta_{XY} \Bigr).
\end{equation}
Since the classical path of Pauli channels~\eqref{eq:PauliGenClass} outputs again a Pauli channel and, therefore, is Choi-stretchable, we can establish an upper bound for its capacity in a similar way as done for the quantum path in Proposition~\ref{prop:2} when it cannot be found as a closed expression via the bottleneck inequality~\eqref{eq:9}. It can be written in the following form,
\begin{eqnarray}
    \label{Eq:34} \nonumber \mathcal{Q}(\mathcal{N}_{\text{C}}) &\leq& 1 + H_2(\tilde{p}_X + \tilde{p}_Y - 2\eta_{XY}) \\
    \nonumber &+& (1 - \tilde{p} + \eta) \log_2 (1 - \tilde{p} + \eta) \\
    \nonumber &+& (\tilde{p}_X - \eta + 2\eta_{YZ})\log_2(\tilde{p}_X - \eta + 2\eta_{YZ}) \\
    \nonumber &+& (\tilde{p}_Y - \eta + 2\eta_{XZ})\log_2(\tilde{p}_Y - \eta + 2\eta_{XZ}) \\
    \nonumber &+& (\tilde{p}_Z - \eta + 2\eta_{XY})\log_2(\tilde{p}_Z - \eta + 2\eta_{XY}) \\
     &\eqdef& \mathcal{Q}_{\text{C}}^{\text{UB}},
\end{eqnarray}
where $H_2(\cdot)$ denotes the binary Shannon entropy. However, it should be noted that this upper bound is, generally speaking, weaker than the upper bound arising from the bottleneck inequality~\eqref{eq:9}.

\begin{rem}
    Although the results derived in Propositions~\ref{prop:3} and \ref{prop:4} bound the capacity of arbitrary Pauli channels through a lower and upper bound  in \eqref{Eq:37} and \eqref{Eq:38} respectively, these bounds depend on several parameters, preventing us from easily assessing the region in which the quantum SWITCH outperforms the limits of conventional quantum Shannon's theory. Hence, in the following subsection, we focus on three classes of Pauli channels with the remarkable property of allowing heralded noiseless communications.
\end{rem}

\subsection{Heralded Noiseless Pauli Channels}
\label{sec:5.2}

Interestingly, for certain Pauli channels $\mathcal{D}$ and $\mathcal{E}$, at least one of the components in equations \eqref{eq:PauliGenCompMinus} and \eqref{eq:PauliGenCompPlus} of the equivalent quantum SWITCH channel can have maximal quantum capacity, hence, enabling perfect communication similar to the known case of identical Pauli channels, $\mathcal{D} = \mathcal{E}$~\cite{ChiBanSan-18}. Indeed, there can be highlighted three \textit{non-trivial} classes (summarized in Table~\ref{Tab:IdealChan}) of channel pairs $(\mathcal{D}, \mathcal{E})$ that activate noiseless (i.e., perfect) communication in one or both components of the equivalent channel:
\begin{itemize}
    \item Class I: channel pairs with one of the components $p_{X, Y, Z}$ being zero in both the channels, i,.e., $p_\alpha^{\mathcal{E}} = p_\alpha^{\mathcal{D}} = 0$ for a certain $\alpha \in {X,Y,Z}$ , making $\mathcal{N}_{\text{QS}}^{\ket{-}}$ unitary, hence, ideal,

    \item Class II: channel pairs with $p_0^{\mathcal{E}} = p_0^{\mathcal{D}} = 0$, making $\mathcal{N}_{\text{QS}}^{\ket{+}}$ identity channel,

    \item Class III: channel pairs where one of the channels is unitary, i.e., one of the components $p_{X, Y, Z}$ being one and the rest being zero, whereas the corresponding component $p_{X, Y, Z}$ of the other channel is zero, making $\mathcal{N}_{\text{QS}}^{\ket{+}}$ unitary.
\end{itemize}
The corresponding components of the equivalent channels and upper bounds on their quantum capacity are summarized in Table~\ref{Tab:IdealChan} for all three classes of Pauli channels.

\begin{rem}
A particular case is constituted by Pauli channels that belong to both classes I and II, i.e., a pair of Pauli channels $(\mathcal{D}, \mathcal{E})$ with $p_0^{\mathcal{D}} = p_0^{\mathcal{E}} = 0$ and $p_\alpha^{\mathcal{D}} = p_\alpha^{\mathcal{E}} = 0$ for a certain $\alpha = \{ X, Y, Z\}$, for example,
\begin{eqnarray}
    \nonumber \tilde{\mathcal{D}} &=& (0, p_X^{\mathcal{D}}, 0, p_Z^{\mathcal{D}}), \\
    \nonumber \tilde{\mathcal{E}} &=& (0, p_X^{\mathcal{E}}, 0, p_Z^{\mathcal{E}}),
\end{eqnarray}
where $\alpha = Y$ without loss of generality. Such channels make both components~\eqref{eq:PauliGenCompMinus} and~\eqref{eq:PauliGenCompPlus} ideal (unitary and identity, respectively). Hence, in this case, the overall equivalent channel becomes perfect~\cite{ChiBanSan-18}, assuring noiseless communications for any measurement of the control qubit.
\end{rem}

\begin{rem}\label{rem:EntBreak} An important subclass of Pauli channels are entanglement-breaking channels, which have zero coherent information and, therefore, zero quantum capacity. When one of the channels put into a classical path is entanglement-breaking, then the quantum capacity of the entire classical path is zero due to the bottleneck inequality~\eqref{eq:9}. Given a Pauli channel with the probability vector $(p_0, p_X, p_Y, p_Z)$, it is entanglement-breaking if~\cite{Ruskai2003, Siudzinska2019}
\begin{equation}\label{eq:EntBreak}
    \sum_{\alpha = X, Y, Z} |2(p_0 + p_\alpha) - 1| \leq 1.
\end{equation}
For the channels $\mathcal{D}$ and $\mathcal{E}$ that make noiseless at least one of the components $\mathcal{N}_{\text{QS}}^{\ket{\mp}}(\cdot)$ of the quantum SWITCH equivalent channel, condition \eqref{eq:EntBreak} means that $\tilde{\mathcal{D}}$ or $\tilde{\mathcal{E}}$ have to satisfy the conditions 
\begin{eqnarray}
p_X^{\mathcal{D}/\mathcal{E}} &\leq& \frac{1}{2}, \\
\frac{1}{2} - p_X^{\mathcal{D}/\mathcal{E}} \leq p_Z^{\mathcal{D}/\mathcal{E}} &\leq& \frac{1}{2},
\end{eqnarray}
in order to be entanglement-breaking. A classical path of such channels has zero capacity, $\mathcal{Q}(\mathcal{N}_{\text{C}}) = 0$.
\end{rem}

\begin{table*}[t]
	\centering
    \begin{tabular}{| p{0.14\textwidth} | p{0.25\textwidth} | p{0.29\textwidth}| p{0.26\textwidth}|}
		\toprule
		\textbf{Parameters} & \textbf{Channel Pair} & \textbf{Equivalent Channel} & \textbf{Noiseless Channel Component}\\
		\midrule
		\textbf{Class I} & \multicolumn{2}{c|}{by assuming $\alpha = Y$ without restriction of generality, it results:} &  $\mathcal{N}_{\text{QS}}^{\ket{-}}$ is unitary \\
		\begin{equation}\nonumber p_\alpha^{\mathcal{D}} = p_\alpha^{\mathcal{E}} = 0\end{equation} & 
		\begin{eqnarray}
        \nonumber \tilde{\mathcal{D}} &=& \Bigl(1 - p_X^{\mathcal{D}} - p_Z^{\mathcal{D}}, p_X^{\mathcal{D}}, 0, p_Z^{\mathcal{D}}\Bigr), \\
        \nonumber \tilde{\mathcal{E}} &=& \Bigl(1 - p_X^{\mathcal{E}} - p_Z^{\mathcal{E}}, p_X^{\mathcal{E}}, 0, p_Z^{\mathcal{E}}\Bigr),
    \end{eqnarray} & \begin{eqnarray}
\nonumber \tilde{\mathcal{N}}_{\text{QS}}^{\ket{-}} &=& \Bigl(0, 0, 1, 0 \Bigr), \\
\nonumber \tilde{\mathcal{N}}_{\text{QS}}^{\ket{+}} &=& \frac{1}{p_+} \Bigl(\sum_{\alpha} p_\alpha^{\mathcal{D}} p_\alpha^{\mathcal{E}}, \eta_{0X}, 0, \eta_{0Z} \Bigr), \\
\nonumber p_- &=& \eta_{XZ}, \\
\nonumber p_+ &=& 1 - \eta_{XZ},
\end{eqnarray} &  \\
        \midrule
        \textbf{Class II} \begin{equation}\nonumber p_0^{\mathcal{D}} = p_0^{\mathcal{E}} = 0\end{equation} & \begin{eqnarray}
        \nonumber \tilde{\mathcal{D}} &=& \Bigl(0, p_X^{\mathcal{D}}, 1 - p_X^{\mathcal{D}} - p_Z^{\mathcal{D}}, p_Z^{\mathcal{D}}\Bigr), \\
        \nonumber \tilde{\mathcal{E}} &=& \Bigl(0, p_X^{\mathcal{E}}, 1 - p_X^{\mathcal{E}} - p_Z^{\mathcal{E}}, p_Z^{\mathcal{E}}\Bigr),
    \end{eqnarray} & \begin{eqnarray}
\nonumber \tilde{\mathcal{N}}_{\text{QS}}^{\ket{-}} &=& \frac{1}{p_-}\Bigl(0, \eta_{YZ}, \eta_{XZ}, \eta_{XY} \Bigr), \\
\nonumber \tilde{\mathcal{N}}_{\text{QS}}^{\ket{+}} &=& \Bigl(1, 0, 0, 0 \Bigr), \\
\nonumber p_- &=& 1 - \sum_{\alpha} p_\alpha^{\mathcal{D}} p_\alpha^{\mathcal{E}}, \\
\nonumber p_+ &=& \sum_{\alpha} p_\alpha^{\mathcal{D}} p_\alpha^{\mathcal{E}}
\end{eqnarray} & $\mathcal{N}_{\text{QS}}^{\ket{+}}$ is identity\\
        \midrule
        \textbf{Class III} & \multicolumn{2}{c|}{by assuming $\alpha = Y$ without restriction of generality, it results:} &  $\mathcal{N}_{\text{QS}}^{\ket{+}}$ is unitary \\
		\begin{equation}\nonumber p_{\alpha}^{\mathcal{D}} = 1
        , \; p_{\alpha}^{\mathcal{E}} = 0\end{equation} & \begin{eqnarray}
        \nonumber \tilde{\mathcal{D}} &=& \Bigl(0, 0, 1, 0\Bigr), \\
        \nonumber \tilde{\mathcal{E}} &=& \Bigl(1 - p_X - p_Z, p_X, 0, p_Z\Bigr).
    \end{eqnarray} & \begin{eqnarray}
\nonumber \tilde{\mathcal{N}}_{\text{QS}}^{\ket{-}} &=& \frac{1}{p_-}\Bigl(0, p_Z, 0, p_X \Bigr), \\
\nonumber \tilde{\mathcal{N}}_{\text{QS}}^{\ket{+}} &=& \Bigl(0, 0, 1, 0 \Bigr), \\
\nonumber p_- &=& p_X + p_Z, \\
\nonumber p_+ &=& 1 - p_X - p_Z,
\end{eqnarray} &  \\
		\bottomrule
	\end{tabular}
	\caption{Three classes of Pauli channel pairs $(\mathcal{D}, \mathcal{E})$ that make ideal at least one component of the equivalent quantum SWITCH channel $\mathcal{N}_{\text{QS}}(\cdot)$.}
	\label{Tab:IdealChan}
	\hrulefill
\end{table*}

\vspace{6pt}
\subsubsection{Class I}
In the following, we assume without lack of generality that $p_\alpha^{\mathcal{E}} = p_\alpha^{\mathcal{D}} = 0$ for $\alpha = Y$. Accordingly, whenever the measurement of the control qubit in the Hadamard basis returns $\ket{-}$, which happens with probability $p_- = \eta_{XZ}$, the original quantum state $\rho$ appears to be transformed unitarily as $Y\rho Y$. Hence, it can be recovered by applying the unitary corrective operation $Y$ to the received qubit, i.e., $Y(Y \rho Y)Y=\rho$. As a consequence, when the measurement outcome is $\ket{-}$, despite $\mathcal{D}$ and $\mathcal{E}$ being noisy channels corrupting the quantum information embedded in $\ket{\varphi}$, a quantum path implemented via the quantum SWITCH allows a noiseless quantum transmission with a probability equal to $\eta_{XZ}$, and the receiver can easily recognize this event by simply measuring the control qubit. Conversely, whenever the measurement outcome of the control qubit is equal to $\ket{+}$, which happens with probability $p_+ = 1 - \eta_{XZ}$, the original quantum state $\rho$ is altered through a weighted combination of bit-flip $X$ and phase-flip $Z$.

\begin{cor}
	\label{prop:1ClassI}
	The quantum capacity $\mathcal{Q}(\mathcal{N}_{\text{\rm QS}})$ of the equivalent quantum SWITCH channel $\mathcal{N}_{\text{\rm QS}}(\cdot)$ for Pauli channels of class I is lower-bounded as follows:
	\begin{eqnarray}
    \nonumber\mathcal{Q}_{\text{\rm QS}}^{\text{\rm LB}} &=& \eta + \max\{ 0,  (1-\eta)(1-\log_2(1-\eta)) \\
    \nonumber &+& (\tilde{p}_X - \eta)\log_2(\tilde{p}_X - \eta) \\
    \nonumber &+& (\tilde{p}_Z - \eta)\log_2(\tilde{p}_Z - \eta) \\
    \nonumber &+& (1-\tilde{p}_X-\tilde{p}_Z+\eta) \\
    &\cdot&\log_2((1-\tilde{p}_X-\tilde{p}_Z+\eta))\},
    \end{eqnarray}
	\begin{IEEEproof}
		By accounting for the hypothesis of $p_Y^\mathcal{D} = p_Y^\mathcal{E} = 0$ without loss of generality, we obtain equations $\eta_{XY} = \eta_{YZ} = 0$, $\eta = \eta_{XZ}$, $\tilde{p}_Y = 0$, and $\tilde{p} = \tilde{p}_X + \tilde{p}_Z$. By plugging them into~\eqref{Eq:37} and by performing some algebraic manipulations, the proof follows.
	\end{IEEEproof}
\end{cor}
\begin{cor}
	\label{prop:2ClassI}
	The quantum capacity $\mathcal{Q}(\mathcal{N}_{\text{\rm QS}})$ of the equivalent quantum SWITCH channel $\mathcal{N}_{\text{\rm QS}}(\cdot)$ for Pauli channels of class I is upper-bounded as follows:
\begin{eqnarray}
    \nonumber \mathcal{Q}_{\text{\rm QS}}^{\text{\rm UB}} &=& 1 + (1-\tilde{p}_X-\tilde{p}_Z+\eta) \\
    \nonumber &\cdot& \log_2((1-\tilde{p}_X-\tilde{p}_Z+\eta)) \\
    \nonumber &-& (1 - \tilde{p}_X)\log_2(1 - \tilde{p}_X) \\
    &+& (\tilde{p}_Z - \eta)\log_2(\tilde{p}_Z - \eta),
\end{eqnarray}
	\begin{IEEEproof}
		By accounting for the hypothesis of $p_Y^\mathcal{D} = p_Y^\mathcal{E} = 0$ without loss of generality, we obtain equations $\eta_{XY} = \eta_{YZ} = 0$, $\eta = \eta_{XZ}$, $\tilde{p}_Y = 0$, and $\tilde{p} = \tilde{p}_X + \tilde{p}_Z$. By plugging them into~\eqref{Eq:38}, and by performing some algebraic manipulations, the proof follows.
	\end{IEEEproof}
\end{cor}

To properly quantify the gain in terms of quantum capacity assured by the adoption of the quantum SWITCH, we report the difference $\delta\mathcal{Q}_{\text{QS/C}} \eqdef \mathcal{Q}_{\text{QS}}^{\text{LB}} - \mathcal{Q}_{\text{C}}^{\text{UB}}$ between the quantum SWITCH lower bound given in Corollary~\ref{prop:1ClassI} and the upper bound on the capacity achievable with a classical path and
given in \eqref{Eq:34}. For Class I, \eqref{Eq:34} simplifies as follows:
\begin{eqnarray}\label{eq:ClassCI}
    \nonumber \mathcal{Q}_{\text{C}}^{\text{UB}} &=& 1 + H_2(\tilde{p}_X) + \eta \log_2 \eta \\
    \nonumber &+& (\tilde{p}_X - \eta)\log_2(\tilde{p}_X - \eta) \\
    \nonumber &+& (\tilde{p}_Z - \eta)\log_2(\tilde{p}_Z - \eta) \\
    \nonumber &+& (1 - \tilde{p}_X - \tilde{p}_Z + \eta) \\
    &\cdot& \log_2(1 - \tilde{p}_X - \tilde{p}_Z + \eta).
\end{eqnarray}
The difference $\delta\mathcal{Q}_{\text{QS/C}}$ represents a conservative estimate of the capacity gain provided by a quantum path, since: i) the quantum SWITCH lower bound underestimates the quantum capacity of the equivalent quantum SWITCH channel, and simultaneously ii) the classical path upper bound overestimates the capacity achievable via a classical path. For the Pauli channels of class I, $\delta\mathcal{Q}_{\text{QS/C}}$ is given by:
\begin{eqnarray}
    \nonumber \delta\mathcal{Q}_{\text{QS/C}} &=& H_2(\eta) - H_2(\tilde{p}_X) \\
    \nonumber &-& \min\Bigl\{ 0, (1-\tilde{p}_X-\tilde{p}_Z+\eta) \\
    \nonumber &\cdot&  \log_2(1-\tilde{p}_X-\tilde{p}_Z+\eta) \\
    \nonumber  &+& (\tilde{p}_X - \eta)\log_2(\tilde{p}_X - \eta) \\
    \nonumber &+& (\tilde{p}_Z - \eta)\log_2(\tilde{p}_Z - \eta) \\
    \nonumber  &+& (1-\eta)(1-\log_2(1-\eta))\Bigr\} \\
    \nonumber &=& H_2(\eta) - H_2(\tilde{p}_X) \\
     &-&\min\{0, p_+ I_c(\mathcal{N}_{\text{QS}}^{\ket{+}})\}.
\end{eqnarray}
Therefore, if coherent information of the non-ideal component of the quantum SWITCH equivalent channel is positive, the lower bound for capacity of quantum path exceeds the upper bound for capacity of classical path if $H_2(\eta) > H_2(\tilde{p}_X)$. Otherwise, the condition reads $p_+ |I_c(\mathcal{N}_{\text{QS}}^{\ket{+}})| > H_2(\tilde{p}_X) - H_2(\eta)$. For the Pauli channels satisfying these conditions, the quantum path necessarily outperforms the corresponding classical path in terms of their quantum capacities.

\vspace{6pt}
\subsubsection{Class II}

Herein, if a measurement of the control qubit in the Hadamard basis returns $\ket{+}$, which happens with probability $p_+ = p_X^\mathcal{D} p_X^\mathcal{E} + p_Y^\mathcal{D} p_Y^\mathcal{E} + p_Z^\mathcal{D} p_Z^\mathcal{E}$, the original quantum state $\rho$ is transmitted perfectly. In this case, we can obtain the following lower and upper bound for quantum capacity of the quantum path.
\begin{cor}
	\label{prop:1ClassII}
	The quantum capacity $\mathcal{Q}(\mathcal{N}_{\text{\rm QS}})$ of the equivalent quantum SWITCH channel $\mathcal{N}_{\text{\rm QS}}(\cdot)$ for Pauli channels of class II is lower-bounded as follows:
	\begin{eqnarray}
    \nonumber \mathcal{Q}_{\text{\rm QS}}^{\text{\rm LB}} &=& 1 - \tilde{p}_X - \tilde{p}_Z + \eta_{XZ} \\
    \nonumber &+& \max\{ 0, \eta_{XZ}\log_2\eta_{XZ} \\
    \nonumber &+& (\tilde{p}_X - \eta_{XZ})\log_2(\tilde{p}_X - \eta_{XZ}) \\
     \nonumber &+& (\tilde{p}_Z - \eta_{XZ})\log_2(\tilde{p}_Z - \eta_{XZ}) \\
     \nonumber &+& (\tilde{p}_X + \tilde{p}_Z - \eta_{XZ}) \\
     &\cdot& (1 - \log_2(\tilde{p}_X + \tilde{p}_Z - \eta_{XZ})) \}.
    \end{eqnarray}
	\begin{IEEEproof}
		By accounting for the hypothesis of $p_0^\mathcal{D} = p_0^\mathcal{E} = 0$ without loss of generality, we obtain equations $\eta_{XY} = \tilde{p}_X - \eta_{XZ}$, $\eta_{YZ} = \tilde{p}_Z - \eta_{XZ}$, $\eta = \tilde{p}_X + \tilde{p}_Z - \eta_{XZ}$, $\tilde{p}_Y = \tilde{p}_X + \tilde{p}_Z - 2\eta_{XZ}$, and $\tilde{p} = 2(\tilde{p}_X + \tilde{p}_Z - \eta_{XZ})$. By plugging them into~\eqref{Eq:37} and by performing some algebraic manipulations, the proof follows.
	\end{IEEEproof}
\end{cor}
\begin{cor}
	\label{prop:2ClassII}
	The quantum capacity $\mathcal{Q}(\mathcal{N}_{\text{\rm QS}})$ of the equivalent quantum SWITCH channel $\mathcal{N}_{\text{\rm QS}}(\cdot)$ for Pauli channels of class II is upper-bounded as follows:
\begin{eqnarray}
    \nonumber \mathcal{Q}_{\text{\rm QS}}^{\text{\rm UB}} &=& 1 + \eta_{XZ}\log_2\eta_{XZ} - \tilde{p}_Z \log_2\tilde{p}_Z \\
    &+& (\tilde{p}_Z - \eta_{XZ}) \log_2(\tilde{p}_Z - \eta_{XZ}).
\end{eqnarray}
\begin{IEEEproof}
	By accounting for the hypothesis of $p_0^\mathcal{D} = p_0^\mathcal{E} = 0$ without loss of generality, we obtain equations $\eta_{XY} = \tilde{p}_X - \eta_{XZ}$, $\eta_{YZ} = \tilde{p}_Z - \eta_{XZ}$, $\eta = \tilde{p}_X + \tilde{p}_Z - \eta_{XZ}$, $\tilde{p}_Y = \tilde{p}_X + \tilde{p}_Z - 2\eta_{XZ}$, and $\tilde{p} = 2(\tilde{p}_X + \tilde{p}_Z - \eta_{XZ})$. By plugging them into~\eqref{Eq:38} and by performing some algebraic manipulations, the proof follows.
\end{IEEEproof}
\end{cor}
By noting that \eqref{Eq:34} simplifies as:
\begin{eqnarray}
    \nonumber \mathcal{Q}_{\text{C}}^{\text{UB}} &=& 1 + H_2(\tilde{p}_Z) + \eta_{XZ} \log_2 \eta_{XZ} \\
    \nonumber &+& (\tilde{p}_X - \eta_{XZ})\log_2(\tilde{p}_X - \eta_{XZ}) \\
    \nonumber &+& (\tilde{p}_Z - \eta_{XZ})\log_2(\tilde{p}_Z - \eta_{XZ}) \\
    \nonumber &+& (1 - \tilde{p}_X - \tilde{p}_Z + \eta_{XZ}) \\
    &\cdot&\log_2(1 - \tilde{p}_X - \tilde{p}_Z + \eta_{XZ}),
\end{eqnarray}
the distance between the upper bound for the capacity of a classical path and the lower bound for the capacity of a quantum path is given by:
\begin{eqnarray}
    \nonumber \delta\mathcal{Q}_{\text{QS/C}} &=& H_2(\tilde{p}_X + \tilde{p}_Z - \eta_{XZ}) - H_2(\tilde{p}_X) \\
    \nonumber &-& \min\Bigl\{ 0, \eta_{XZ}\log_2\eta_{XZ} \\
    \nonumber &+& (\tilde{p}_X - \eta_{XZ})\log_2(\tilde{p}_X - \eta_{XZ}) \\
    \nonumber &+& (\tilde{p}_Z - \eta_{XZ})\log_2(\tilde{p}_Z - \eta_{XZ}) \\
    \nonumber &+& (\tilde{p}_X + \tilde{p}_Z - \eta_{XZ}) \\
    \nonumber &\cdot& (1 - \log_2(\tilde{p}_X + \tilde{p}_Z - \eta_{XZ})) \Bigr\} \\
    \nonumber &=& H_2(\tilde{p}_X + \tilde{p}_Z - \eta_{XZ}) - H_2(\tilde{p}_X) \\
    &-& \min\{0, p_- I_c(\mathcal{N}_{\text{QS}}^{\ket{-}})\}.
\end{eqnarray}
Therefore, if coherent information of the non-ideal component of the quantum SWITCH equivalent channel is positive, the lower bound for capacity of quantum path exceeds the upper bound for capacity of classical path if $H_2(\tilde{p}_X + \tilde{p}_Z - \eta) > H_2(\tilde{p}_X)$. Otherwise, the condition reads $p_- |I_c(\mathcal{N}_{\text{QS}}^{\ket{-}})| > H_2(\tilde{p}_X) - H_2(\tilde{p}_X + \tilde{p}_Z - \eta)$.

\vspace{6pt}
\subsubsection{Class III}

For the third class, without restriction of generality, $p_Y^\mathcal{D} = 1$ (hence, the rest of components are zero). By reasoning as in the previous Corollaries, the lower and upper bounds reported in Corollary~\ref{prop:1ClassIII} and \ref{prop:2ClassIII} can be derived.

\begin{cor}
\label{prop:1ClassIII}
    The quantum capacity $\mathcal{Q}(\mathcal{N}_{\text{\rm QS}})$ of the equivalent quantum SWITCH channel $\mathcal{N}_{\text{\rm QS}}(\cdot)$ for Pauli channels of class III is lower-bounded as follows:
	\begin{eqnarray}\label{eq:ClassIIICap}
    \nonumber \mathcal{Q}_{\text{\rm QS}}^{\text{\rm LB}} &=& 1 + p_X \log_2 p_X + p_Z \log_2 p_Z \\
    &-& (p_X + p_Z) \log_2 (p_X + p_Z) .
\end{eqnarray}
\begin{IEEEproof}
        By adopting similar reasoning as in Corollary~\ref{prop:1ClassII} and by noticing that $\eta_{XY} = p_X$, $\eta_{YZ} = p_Z$, $\eta_{XZ} = 0$, $\eta = p_X + p_Z$, $\tilde{p}_X = p_X$, $\tilde{p}_Y = 1$, $\tilde{p}_Z = p_Z$ and $\tilde{p} = 1 + p_X + p_Z$, the proof follows.
\end{IEEEproof}
\end{cor}

\begin{cor}
\label{prop:2ClassIII}
    The quantum capacity $\mathcal{Q}(\mathcal{N}_{\text{\rm QS}})$ of the equivalent quantum SWITCH channel $\mathcal{N}_{\text{\rm QS}}(\cdot)$ for Pauli channels of class III is upper-bounded as follows:
	\begin{equation}
     \mathcal{Q}_{\text{\rm QS}}^{\text{\rm UB}} = 1.
\end{equation}
\begin{IEEEproof}
        By adopting similar reasoning as in Corollary~\ref{prop:2ClassII} and by noticing that $\eta_{XY} = p_X$, $\eta_{YZ} = p_Z$, $\eta_{XZ} = 0$, $\eta = p_X + p_Z$, $\tilde{p}_X = p_X$, $\tilde{p}_Y = 1$, $\tilde{p}_Z = p_Z$ and $\tilde{p} = 1 + p_X + p_Z$, the proof follows.
\end{IEEEproof}
\end{cor}

\begin{figure}[t]
    \centering
	\includegraphics[width=\columnwidth]{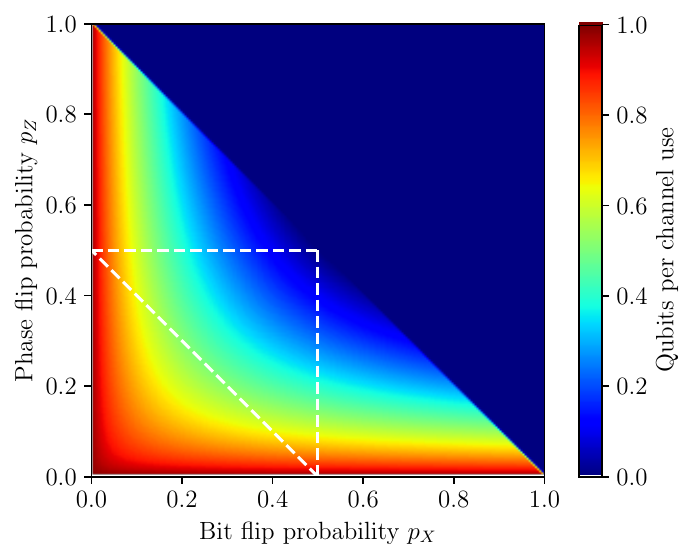}
	\caption{Density plot of the lower bound \eqref{eq:ClassIIICap} on quantum capacity of the quantum path for Pauli channels of class III. The white triangle highlights the area of probabilities which makes one the corresponding Pauli channel entanglement-breaking and, due to Remark~\ref{rem:EntBreak}, quantum capacity of the corresponding classical path can be calculated explicitly, $\mathcal{Q}(\mathcal{N}_{\text{C}}) = 0$. Notice that due to the condition $p_X + p_Z \leq 1$ of consistency of probabilities only the lower half of the plot is considered.}
	\label{Fig:AIII}
	\hrulefill
\end{figure}

The lower bound on the quantum capacity achievable through a quantum path for Pauli channels of class III as a function of $p_X$ and $p_Z$ is plotted on Fig.~\ref{Fig:AIII}.

By noting that \eqref{Eq:34} simplifies as:
\begin{eqnarray}
    \nonumber \mathcal{Q}_{\text{C}}^{\text{UB}} &=& 1 + H_2(p_X) \\
    \nonumber &+& p_X \log_2 p_X + p_Z \log_2 p_Z \\
     &+& (1 - p_X - p_Z) \log_2(1 - p_X - p_Z),
\end{eqnarray}
the distance between the upper bound for the capacity of classical path and the lower bound for the capacity of quantum path is given by
\begin{equation}
    \delta\mathcal{Q}_{\text{QS/C}} = H_2(p_X + p_Z) - H_2(p_X).
\end{equation}

\subsection{Bit- and phase-flip channels}
\label{sec:3C}

In this subsection, we consider a particular subclass of class I, namely Pauli channels $\mathcal{P}$ that allow one to calculate their quantum capacity $\mathcal{Q}(\mathcal{P})$ explicitly. For the sake of simplicity, we take\footnote{This choice is not restrictive, since other types of depolarizing channels are unitarily equivalent to a combination of bit-flip and phase-flip channels \cite{SalEblChi-18}. Hence, the analysis in the following continues to hold and it can be generalized to any depolarizing channel by exploiting proper pre- and post-processing operations.} channel $\mathcal{D}$ being the \textit{bit-flip channel} and channel $\mathcal{E}$ being the \textit{phase-flip channel} as in \cite{SalEblChi-18,CalCac-19}, whose probability vectors are given by:

\begin{align}
	\label{eq:10}
	&\tilde{\mathcal{D}} = \Bigl( 1-p, p, 0, 0 \Bigr) \\
	\label{eq:11}
	&\tilde{\mathcal{E}} = \Bigl( 1-q, 0, 0, q \Bigr),
\end{align}
From \eqref{eq:10}, it appears clear that channel $\mathcal{D}$ flips the state of a qubit from $\ket{0}$ to $\ket{1}$ (and vice versa) with probability $p$, and it leaves the qubit unaltered with probability $1-p$. Similarly, from \eqref{eq:11}, channel $\mathcal{E}$ introduces with probability $q$ a relative phase-shift of $\pi$ between the complex amplitudes $\alpha$ and $\beta$ of the qubit $\ket{\varphi} = \alpha \ket{0} + \beta \ket{1}$, and it leaves the qubit unaltered with probability $1-q$.

\begin{figure}[t!]
	\centering
	\includegraphics[width=\columnwidth]{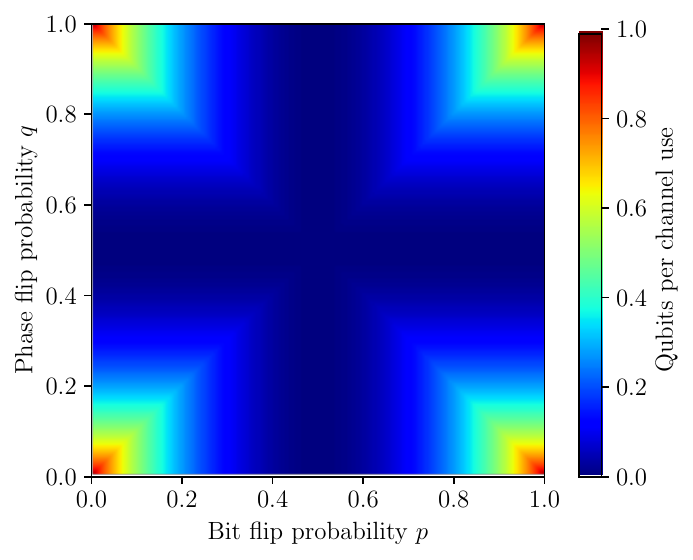}
	\caption{Density plot of the upper bound~\eqref{eq:14} on quantum capacity achievable with a classical path as a function of the error probabilities $p$ and $q$ of bit- and phase-flip channels $\mathcal{D}$ and $\mathcal{E}$. }
	\label{Fig:A2}
	\hrulefill
\end{figure}

Both the bit- and the phase-flip channels admit a single-letter expression for the quantum capacity \cite{Wil-13}:
\begin{align}
	\label{eq:12}
	\mathcal{Q}(\mathcal{D}) &= 1 -H_2(p)\\
	\label{eq:13}
	\mathcal{Q}(\mathcal{E}) &= 1 -H_2(q).
\end{align}
This allows us to substitute the upper bound~\eqref{eq:ClassCI} on the quantum capacity achievable with a classical paths for class I channels, with an improved upper bound arising from the bottleneck inequality in~\eqref{eq:9} for $\mathcal{D}$ and $\mathcal{E}$, namely,
\begin{equation}
	\label{eq:14}
	\mathcal{Q}^{\text{UB}}_{\text{C}} \equiv 1 - \max \left\{ H_2(p),H_2(q) \right\}.
\end{equation}
In Fig.~\ref{Fig:A2} we plot it as a function of the error probabilities $p$ and $q$ of $\mathcal{D}$ and $\mathcal{E}$. 

\begin{rem}
	\label{rem:1}
	Whenever $p$ or $q$ are equal to $\frac{1}{2}$, no quantum information can be sent through any classical path traversing the channels $\mathcal{D}$ and $\mathcal{E}$, since $\mathcal{Q}^{\text{\rm UB}}_{\text{\rm C}} = 0$.
\end{rem}

According to Table~\ref{Tab:IdealChan}, a quantum path for $\mathcal{D}$ and $\mathcal{E}$ leads to the following components of the equivalent quantum SWITCH channel \cite{SalEblChi-18,ChiKri-19,CalCac-19}:
\begin{eqnarray}
	\label{eq:17}
	\tilde{\mathcal{N}}_{\text{QS}}^{\ket{-}} &=& (0, 0, 1, 0) \\ 
	\tilde{\mathcal{N}}_{\text{QS}}^{\ket{+}} &=& \frac{1}{pq}\Bigl( (1-p)(1-q), p (1-q), 0, (1-p) q \Bigr), \label{eq:17a}
\end{eqnarray}
with measurement outcome probabilities $p_- = pq$ and $p_+ = 1 - pq$, respectively. Corollaries~\ref{prop:1ClassI} and \ref{prop:2ClassI} allow one to derive from~\eqref{eq:17} and~\eqref{eq:17a} the lower and upper bounds on the capacity achievable with a quantum path for $\mathcal{D}$ and $\mathcal{E}$.

\begin{figure}[t!]
	\centering
	\includegraphics[width=\columnwidth]{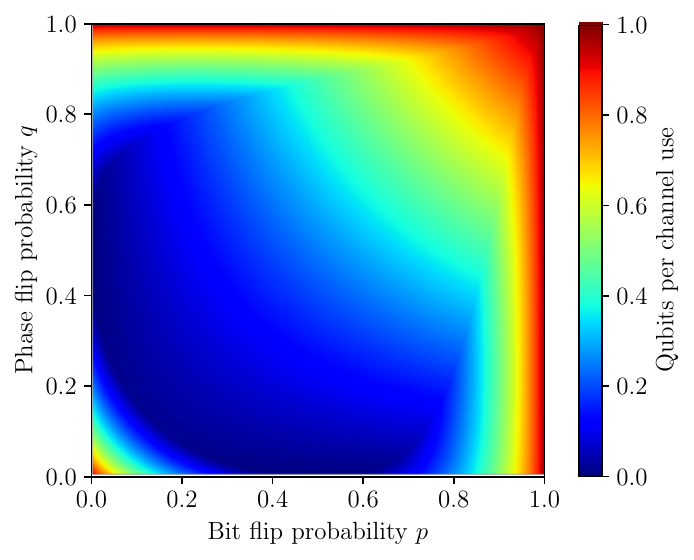}
	\caption{Density plot of the lower bound~\eqref{eq:20} on quantum capacity achievable with the quantum SWITCH as a function of the error probabilities $p$ and $q$ of bit- and phase-flip channels $\mathcal{D}$ and $\mathcal{E}$.}
	\label{Fig:A3}
	\hrulefill
\end{figure}

\begin{cor}
	\label{prop:9}
	The quantum capacity $\mathcal{Q}(\mathcal{N}_{\text{\rm QS}})$ of the equivalent quantum SWITCH channel $\mathcal{N}_{\text{\rm QS}}(\cdot)$ for bit- and phase-flip channels is lower-bounded as follows:
	\begin{eqnarray}		
		\label{eq:20}
		\nonumber \mathcal{Q}_{\text{\rm QS}}^{\text{\rm LB}} &= & \, pq + \max\{0, 1-pq + H_2(pq)-H_2(p)-H_2(q)\} \\
		& =& \begin{cases}
			pq, \; \text{if } 1-pq + H_2(pq)-H_2(p)-H_2(q) <0\\
			1+ H_2(pq)-H_2(p)-H_2(q), \; \text{otherwise}.
		\end{cases}
	\end{eqnarray}	
	\begin{IEEEproof}
		By accounting for the probability vectors given in \eqref{eq:10} and \eqref{eq:11}, we obtain the equations $\eta_{XY} = \eta_{YZ} = 0$, $\eta_{XZ} = \eta = pq$, $\tilde{p}_X = p$, $\tilde{p}_Y = 0$, $\tilde{p}_Z = q$, and $\tilde{p} = p + q$. By plugging them into~\eqref{Eq:37} and performing some algebraic manipultions, the proof follows.
	\end{IEEEproof}
\end{cor}
\begin{cor}
	\label{prop:10}
	The quantum capacity $\mathcal{Q}(\mathcal{N}_{\text{\rm QS}})$ of the equivalent quantum SWITCH channel $\mathcal{N}_{\text{\rm QS}}(\cdot)$ for bit- and phase-flip channels is upper-bounded as follows:
	\begin{equation}
		\label{eq:21}
		\mathcal{Q}^{\text{\rm UB}}_{\text{\rm QS}} = 1-(1-p)H_2(q)
	\end{equation}
	\begin{IEEEproof}
		By accounting for the probability vectors given in \eqref{eq:10} and \eqref{eq:11}, we obtain the equations $\eta_{XY} = \eta_{YZ} = 0$, $\eta_{XZ} = \eta = pq$, $\tilde{p}_X = p$, $\tilde{p}_Y = 0$, $\tilde{p}_Z = q$, and $\tilde{p} = p + q$. By plugging them in to~\eqref{Eq:38} and performing some algebraic manipultions, the proof follows.
	\end{IEEEproof}
\end{cor}

\begin{figure}[t]
	\centering
	\includegraphics[width=\columnwidth]{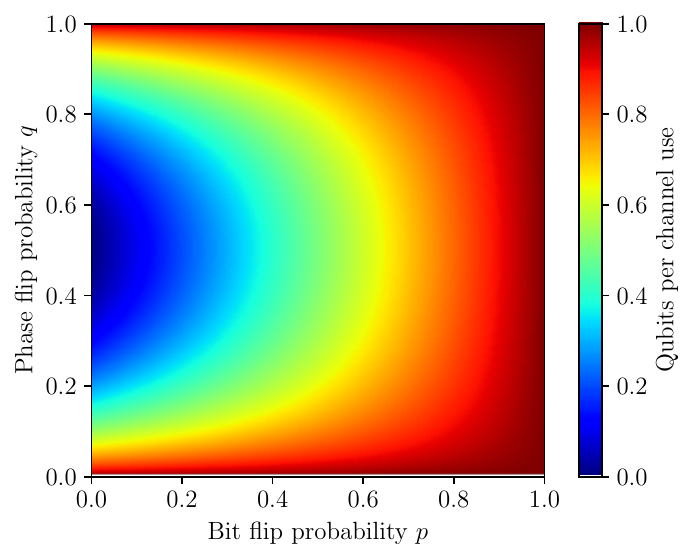}
	\caption{Density plot of the upper bound~\eqref{eq:21} on  quantum capacity achievable with the quantum SWITCH as a function of the error probabilities $p$ and $q$ of bit- and phase-flip channels $\mathcal{D}$ and $\mathcal{E}$. }
	\label{Fig:A4}
	\hrulefill
\end{figure}

In Fig.~\ref{Fig:A3} and Fig.~\ref{Fig:A4} we plot the lower bound~\eqref{eq:20} and upper bound~\eqref{eq:21}, respectively, as functions of the error probabilities $p$ and $q$ of the channels. 

A particular interest is brought by $\mathcal{D}$ and $\mathcal{E}$, whereof at least one has a null capacity (i.e., $p = \frac{1}{2}$ or $q = \frac{1}{2}$) reducing to zero the capacity of the corresponding classical path in accordance with~\eqref{eq:14}. In this case, for any meaningful choice of $p, q \neq 0$, the quantum SWITCH assures a non-null capacity.
In particular, in the extreme case of $\mathcal{D}$ and $\mathcal{E}$ having both zero capacity, the quantum capacity achievable by utilizing the quantum SWITCH is greater than zero,
\begin{equation}
	\label{eq:21_bis_2}
	\frac{1}{4} \leq \mathcal{Q}(\mathcal{N}_{\text{QS}}) \leq \frac{1}{2}.
\end{equation}

\begin{figure}[t]
	\centering
	\includegraphics[width=\columnwidth]{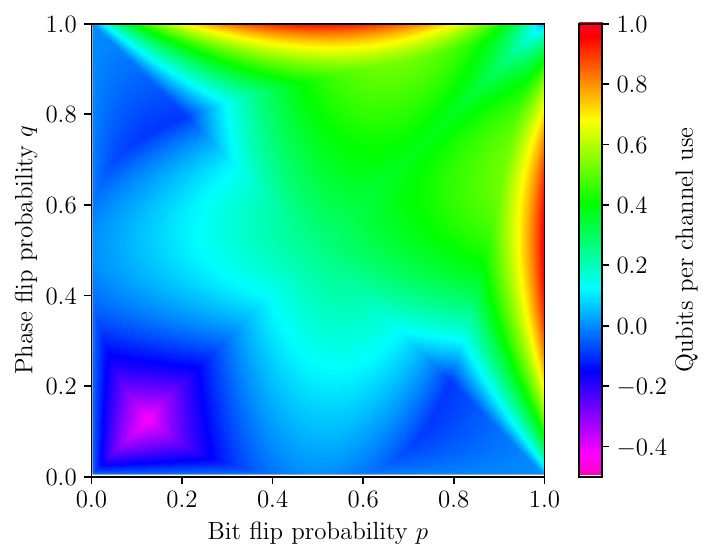}
	\caption{Density plot of the difference between the quantum SWITCH lower bound~\eqref{eq:20} and the classical path upper bound~\eqref{eq:14} as a function of the error probabilities $p$ and $q$ of bit- and phase-flip channels $\mathcal{D}$ and $\mathcal{E}$.}
	\label{Fig:DeltaBP}
	\hrulefill
\end{figure}

To properly quantify the gain in terms of quantum capacity assured by the adoption of the quantum SWITCH, in Fig.~\ref{Fig:DeltaBP} we plot the difference $\delta\mathcal{Q}_{\text{QS/C}}$ between the lower bound $\mathcal{Q}^{\text{LB}}_{\text{QS}}$ for quantum path and the upper bound $\mathcal{Q}^{\text{UB}}_{\text{C}}$ for classical path.
Despite being $\delta\mathcal{Q}_{\text{QS/C}}$ an underestimation of the actual gain provided by a quantum path, we can observe in Fig.~\ref{Fig:DeltaBP} that the adoption of the quantum SWITCH incontrovertibly increases the performance in terms of capacity with respect to classical paths within the broad range of values of the error probabilities $p$ and $q$ that correspond to the region with $\delta\mathcal{Q}_{\text{QS/C}} > 0$.
For any setting of $p$ and $q$ within this wide region, the quantum path enables with certainty the violation of the bottleneck inequality given in \eqref{eq:9}. A particular example is drawn by the extreme case of $\delta\mathcal{Q}_{\text{QS/C}} = 1$, which represents a setting for $p$ and $q$ where the equivalent quantum SWITCH channel behaves as an ideal channel, even if no quantum information at all can be sent through any classical path. Indeed, this is achieved for $p = \frac{1}{2}$ and $q = 1$ or vice versa.

\begin{figure}[t]
	\centering
	\includegraphics[width=\columnwidth]{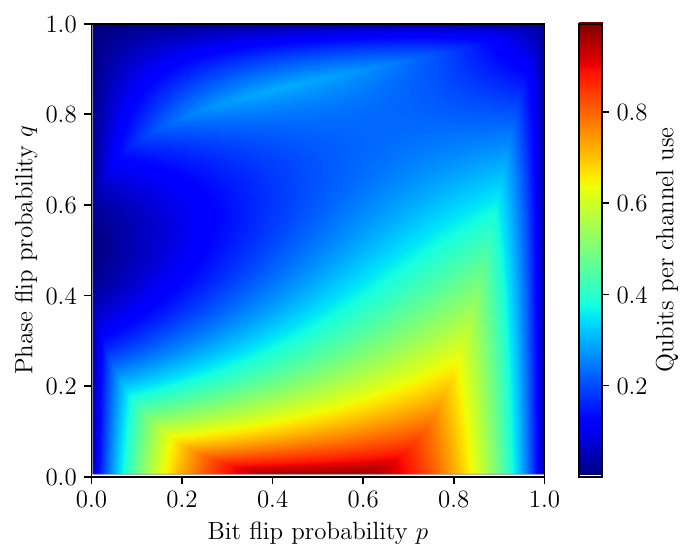}
	\caption{Density plot of the difference between the upper bound~\eqref{eq:21} and lower bound~\eqref{eq:20} on quantum capacity achievable with the quantum SWITCH as a function of the error probabilities $p$ and $q$ of bit- and phase-flip channels $\mathcal{D}$ and $\mathcal{E}$.}
	\label{Fig:A6}
	\hrulefill
\end{figure}

Regarding the regions with negative $\delta\mathcal{Q}_{\text{QS/C}}$, they do not imply that a classical path outperforms the quantum path, given that capacity bounds -- rather than exact capacities -- are involved within $\delta\mathcal{Q}_{\text{QS/C}}$. To further discuss this point, in Fig.~\ref{Fig:A6} we plot the difference $\delta\mathcal{Q}_{\text{QS}} \eqdef \mathcal{Q}^{\text{UB}}_{\text{QS}} - \mathcal{Q}^{\text{LB}}_{\text{QS}}$ between the upper- and lower bounds of the quantum capacity achievable via a quantum path for the bit- and the phase-flip channels $\mathcal{D}$ and $\mathcal{E}$ given in \eqref{eq:10} and \eqref{eq:11}. Roughly speaking, this quantity measures the \textit{uncertainty} about the true value of the quantum SWITCH capacity $\mathcal{Q}(\mathcal{N}_{\text{QS}})$. Fig.~\ref{Fig:A6} highlights that $\delta\mathcal{Q}_{\text{QS}} \lesssim 0.33$ for most of the considered values for the error probabilities $p$ and $q$, indicated with colors ranging from purple to turquoise. This is an indication of a good tightness of the derived bounds. Conversely, as $q$ goes to zero, the difference between the two bounds increases, reaching the maximum value of $1$ qubit when $p=\frac{1}{2}$. Regardless of the tightness quality, the aforementioned analysis still holds, since we compare the lower bound on the quantum path capacity with the upper bound on the classical path capacity. Moreover, this comparison incontrovertibly quantifies the advantages achievable with the quantum SWITCH over any classical path.

Before to close the discussion on bit- and the phase-flip channels, let us focus on the case of $\mathcal{D}$ and $\mathcal{E}$ characterized by the same error probability $p=q$. In particular, this setting allows one a direct comparison with the analysis developed in \cite{SalEblChi-18} for the lower bound on the quantum capacity achievable by utilizing the quantum SWITCH.

\begin{cor}
	\label{cor:1}
	When $p=q$, the lower bound $\mathcal{Q}^{\text{\rm LB}}_{\text{\rm QS}}$ in \eqref{eq:20} on the quantum capacity achievable via a quantum path implemented through the quantum SWITCH can be rewritten as:
	\begin{align}
		\label{eq:23}
		\mathcal{Q}^{\text{\rm LB}}_{\text{\rm QS}} = \begin{cases}
				p^2, & \text{for } p \in [p_0, 1] \\
				1+ H_2(p^2)-2H_2(p), & \text{otherwise}
			\end{cases}
	\end{align}
	with $p_0 \simeq 0.128$.
	\begin{IEEEproof}
		For $p=q$, the argument within the maximum operator in \eqref{eq:20} reduces to $1-p^2+H_2(p^2)-2H_2(p)$. By solving the inequality $1-p^2+H_2(p^2)-2H_2(p)<0$, it results that when $p_0 \leq p \leq 1$ the aforementioned inequality is satisfied. Hence, the proof follows.
	\end{IEEEproof}
\end{cor}

\begin{figure}[t]
	\centering
	\includegraphics[width=\columnwidth]{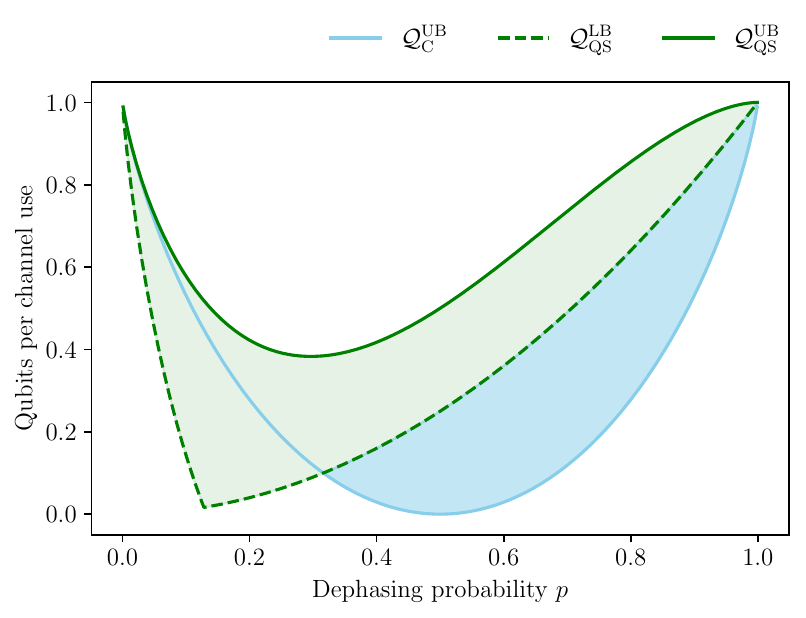}
	\caption{Quantum capacity bounds as a function of the error probabilities $q=p$ of the noisy channels $\mathcal{D}$ and $\mathcal{E}$. The green area denotes the region -- enclosed by the lower- and the upper bound $\mathcal{Q}^{\text{LB}}_{\text{QS}}$ and $\mathcal{Q}^{\text{UB}}_{\text{QS}}$ -- where the quantum SWITCH capacity $\mathcal{Q}(\mathcal{N}_{\text{QS}})$ belongs to, whereas the blue area denotes a conservative estimate of the quantum SWITCH advantage over classical paths.}
	\label{Fig:A1}
    \hrulefill
\end{figure}

By directly comparing \eqref{eq:23} with Eq.~(8) in \cite{SalEblChi-18} (associated Figure~2), it is evident the higher accuracy of the lower bound in \eqref{eq:23}. In Fig.~\ref{Fig:A1} we plot the upper bound $\mathcal{Q}^{\text{UB}}_{\text{C}}$ from a classical path and the lower and upper bounds $\mathcal{Q}^{\text{LB}}_{\text{QS}}$ and $\mathcal{Q}^{\text{UB}}_{\text{QS}}$ from a quantum path. Specifically, we observe that the adoption of the quantum SWITCH incontrovertibly boosts the performance in terms of achievable quantum capacity and allows one to violate the bottleneck inequality~\eqref{eq:9}, whenever $p$ is greater than a threshold value roughly equal to $0.3$, as detailed in Corollary~\ref{cor:2}.

\begin{cor}
	\label{cor:2}
	When $p=q$, we have the following:
	\begin{equation}
		\label{eq:24}
		\mathcal{Q}(\mathcal{N}_{\text{\rm QS}}) > \mathcal{Q}(\mathcal{D} \rightarrow \mathcal{E}) \; \forall \, p \geq p_1
	\end{equation}
	with $p_1 \simeq 0.3161$. Clearly, the same result holds for $\mathcal{Q}(\mathcal{E} \rightarrow \mathcal{D})$.
\begin{IEEEproof}
		The proof follows by accounting for the result in Corollary~\ref{cor:1} and by performing some algebraic manipulations. Specifically, it is possible to recognize that for $p_1 \leq p \leq 1$ the lower bound in \eqref{eq:23} is greater than the upper bound in \eqref{eq:14}: 
		\begin{equation}
		\label{eq:24_bis}
		\mathcal{Q}(\mathcal{D} \rightarrow \mathcal{E}) \leq 1-H_2(p)\leq p^2 \leq \mathcal{Q}(\mathcal{N}_{\text{\rm QS}}), \forall \, p \geq p_1
	\end{equation}
	\end{IEEEproof}
\end{cor}

We note that the advantage of the quantum SWITCH over classical paths was shown to exist in literature for values of $p\geq 0.62$ in \cite{SalEblChi-18}. Corollary~\ref{cor:2} improves this result by extending the range in which the lower bound on the quantum capacity achievable by utilizing the quantum SWITCH exceeds the upper bound on the quantum capacity achievable via classical path. This is due to the higher accuracy of the lower bound derived in this manuscript.

\section{Conclusions}
\label{sec:6} 

In this work, we investigated the ultimate rates achievable with quantum paths implemented via the quantum SWITCH. Specifically, we derived expressions for both the upper- and the lower bound on the quantum capacity for different popular quantum channels. Some of the derived expressions depend, remarkably, on computable single-letter quantities, whereas for same particular cases the upper and the lower bounds coincide. Our findings reveal the substantial advantage achievable with a quantum path over any classical combination of the communications channels in terms of ultimate achievable communication rates. Furthermore, we identified the region where a quantum path incontrovertibly outperforms the amount of transmissible information beyond the limits of conventional quantum Shannon's theory, and we quantified this advantage over classical paths through a conservative estimate.

\appendices

\section{Proof of Proposition~\ref{prop:1}}
\label{App:1}

By accounting for the equivalent channel model presented in Sec.~\ref{sec:3}, let us denote with $\mathcal{Q}(\mathcal{N}_{\text{QS}}^{\ket{-}})$ the quantum capacity of the equivalent quantum SWITCH channel heralded by a $\ket{-}$-measurement of the control qubit $\ket{\varphi_c}$, and with $\mathcal{Q}(\mathcal{N}_{\text{QS}}^{\ket{+}})$ the quantum capacity of the quantum SWITCH channel heralded by a $\ket{+}$-measurement of the control qubit $\ket{\varphi_c}$. Since these two events, occurring with the probabilities $p_-$ and $p_+$ given in \eqref{Eq:17}, are disjoint, by accounting for the total probability theorem it is possible to express the overall quantum capacity $\mathcal{Q}(\mathcal{N}_{\text{QS}})$ of the equivalent quantum SWITCH channel $\mathcal{N}_{\text{QS}}(\cdot)$ as:
\begin{equation}
	\label{eq:app.0.1}
	\mathcal{Q}(\mathcal{N}_{\text{QS}}) = p_- \,\mathcal{Q}(\mathcal{N}^{\ket{-}}_{\text{QS}}) + p_+ \,\mathcal{Q}(\mathcal{N}^{\ket{+}}_{\text{QS}}).
\end{equation}

By accounting for \eqref{eq:8} it results that the channel coherent information constitutes a lower bound -- in the region where it is not negative -- for the quantum capacity $\mathcal{Q}(\mathcal{N}^{\ket{-}}_{\text{QS}})$ of the quantum SWITCH channel\footnote{Any quantum channel $\mathcal{N}$ satisfying $\mathcal{N}(I) = I$, i.e., mapping the identity to itself, is a \textit{unital channel}. For unital channels, the channel coherent information $I_{c}(\mathcal{N})$ defined in Definition~\ref{def:5} and the channel reverse coherent information $I_{rc}(\mathcal{N})$ coincide \cite{PirLauOtt-17}. It is straightforward to verify that the quantum SWITCH channel $\mathcal{N}_{\text{QS}}(\cdot)$ given in \eqref{Eq:19} is a \textit{unital channel} for both the cases of measuring the control qubit either in the state $\ket{-}$ or in the state $\ket{+}$.}  heralded by a $\ket{-}$-measurement, and the same holds for the quantum capacity $\mathcal{Q}(\mathcal{N}^{\ket{+}}_{\text{QS}})$. Hence, it results:
\begin{equation}
	\label{eq:app:3}
	\mathcal{Q}(\mathcal{N}_{\text{QS}}^{\ket{\mp}}) \geq \max\{0,I_{c}(\mathcal{N}_{\text{QS}}^{\ket{\mp}})\},
\end{equation}
where $I_{c}(\mathcal{N})$ is defined in \eqref{eq:7} and the $\max$ operator accounts for the case of negative coherent information. Therefore, the quantum capacity of the overall equivalent channel is lower bounded as:
\begin{align}
	\label{eq:app:0.2}
	\nonumber \mathcal{Q}(\mathcal{N}_{\text{QS}}&) \geq p_- \max\{0,I_{c}(\mathcal{N}_{\text{QS}}^{\ket{-}})\} + p_+ \max\{0,I_{c}(\mathcal{N}_{\text{QS}}^{\ket{+}})\} \\
    =& \; \max\{0, p_- I_{c}(\mathcal{N}_{\text{QS}}^{\ket{-}})\} + \max\{0, p_+ I_{c}(\mathcal{N}_{\text{QS}}^{\ket{+}})\},
\end{align}
where the last equality arises from non-negativity of the probabilities $p_\pm \geq 0$.
\section{Proof of Proposition~\ref{prop:2}}
\label{App:2}

It is known that quantum capacity of a Choi-stretchable channel $\mathcal{N}$ is bounded from above by relative entropy of entanglement $E_R(\rho_{\mathcal{N}})$ of its Choi matrix~\cite{PirLauOtt-17}. Hence, if the components $\mathcal{N}_{\text{QS}}^{\ket{\mp}}(\cdot)$ of the equivalent quantum SWITCH channel are Choi-stretchable, their quantum capacities are upper-bounded as:
\begin{equation}
	\label{eq:app:11}
	\mathcal{Q}(\mathcal{N}_{\text{QS}}^{\ket{\mp}})\leq E_R(\rho_{\mathcal{N}_{\text{QS}}^{\ket{\mp}}}).
\end{equation}
Therefore, since quantum capacity of the equivalent quantum SWITCH channel is the average quantum capacity~\eqref{eq:app.0.1} of its components, it is upper-bounded as:
\begin{equation}
	\label{eq:app:11.1}
	\mathcal{Q}(\mathcal{N}_{\text{QS}})\leq p_- E_R(\rho_{\mathcal{N}_{\text{QS}}^{\ket{-}}}) + p_+ E_R(\rho_{\mathcal{N}_{\text{QS}}^{\ket{+}}}).
\end{equation}

\section{{Proof of Proposition~\ref{prop:3}}}
\label{App:3}

To prove the proposition and accordingly to~\eqref{eq:app:0.2}, it is necessary to calculate channel coherent information $I_c(\mathcal{N}_{\text{QS}}^{\ket{\mp}})$ of both the components of the equivalent quantum SWITCH channel (characterized by their probability vectors~\eqref{eq:PauliGenCompMinus} and~\eqref{eq:PauliGenCompPlus}). Accordingly to~\eqref{eq:4}, we extend the channels $\mathcal{N}_{\text{QS}}^{\ket{\mp}}(\cdot)$ as
\begin{equation}
    \rho_{RB} = (\mathcal{I}_R \otimes \mathcal{N}_{\text{QS}}^{\ket{\mp}})(\rho_{RA})
\end{equation}
With respect to the definition of channel coherent information~\eqref{eq:7} and by reasoning as in \cite{PirLauOtt-17,SalEblChi-18,LedLeuSmi-18}, it can be shown that the state $\ket{\psi_{RA}}$ maximizing~\eqref{eq:7} is the maximally entangled state between systems $A$ and $R$, i.e., $\ket{\psi_{RA}} = \ket{\Phi^+}$. Hence, taking into account the definition of Choi matrix~\eqref{eq:5}, $\rho_{RB}$ is the Choi matrix $\rho_{\mathcal{N}_{\text{QS}}^{\ket{\mp}}}$ of the corresponding channel $\mathcal{N}_{\text{QS}}^{\ket{\mp}}(\cdot)$. Hence, the channel coherent information $I_c(\mathcal{N}_{\text{QS}}^{\ket{\mp}})$ is given by:
\begin{equation}
	\label{eq:app:4}
	I_{c}(\mathcal{N}_{\text{QS}}^{\ket{\mp}}) = S(\rho_B^\mp)-S(\rho_{\mathcal{N}_{\text{QS}}^{\ket{\mp}}}),
\end{equation}
where $\rho_B^\mp = \text{Tr}_R[\rho_{\mathcal{N}_{\text{QS}}^{\ket{\mp}}}]$.
For two not necessarily identical Pauli channels, by taking into account \eqref{eq:PauliGenCompMinus} and~\eqref{eq:PauliGenCompPlus}) and 
\begin{eqnarray}
(I_R \otimes X)\rho_{\Phi^+}(I_R \otimes X) &=& \rho_{\Psi^+}, \\
(I_R \otimes Y)\rho_{\Phi^+}(I_R \otimes Y) &=& \rho_{\Psi^-}, \\
(I_R \otimes Z)\rho_{\Phi^+}(I_R \otimes Z) &=& \rho_{\Phi^-},
\end{eqnarray}
where $\rho_{\Phi^\pm} = \ket{\Phi^\pm} \bra{\Phi^\pm}$ and $\rho_{\Psi^\pm} = \ket{\Psi^\pm} \bra{\Psi^\pm}$ with $\ket{\Phi^\pm} = \frac{1}{\sqrt{2}}(\ket{00} \pm \ket{11})$ and $\ket{\Psi^\pm} = \frac{1}{\sqrt{2}}(\ket{01} \pm \ket{10})$, the corresponding Choi matrices can be given as:
\begin{eqnarray}
    \rho_{\mathcal{N}_{\text{QS}}^{\ket{-}}} &=& \frac{1}{p_-} \Bigl( \eta_{YZ} \rho_{\Psi^+} + \eta_{XZ} \rho_{\Psi^-} + \eta_{XY} \rho_{\Phi^-} \Bigr), \\
    \nonumber \rho_{\mathcal{N}_{\text{QS}}^{\ket{+}}} &=& \frac{1}{p_+} \Bigl( \Bigl( \sum_\alpha \eta_{\alpha\alpha} \Bigr) \rho_{\Phi^+} + \eta_{0X} \rho_{\Psi^+} \\
    &+& \eta_{0Y} \rho_{\Psi^-} +\eta_{0Z} \rho_{\Phi^-} \Bigr),
\end{eqnarray}
with the probabilities
\begin{eqnarray}
    \label{eq:app:pMinus} p_- &=& \eta, \\
    \label{eq:app:pPlus} p_+ &=& 1- \eta.
\end{eqnarray}
The first term in \eqref{eq:app:4} can be evaluated straightforwardly, since $\text{Tr}_R[\rho_{\Phi^\pm}] = \text{Tr}_R[\rho_{\Psi^\pm}] = \frac{\mathbb{I}}{2}$, so that
\begin{equation}
    \rho_B^\mp = \frac{\mathbb{I}}{2},
\end{equation}
where $\mathbb{I}$ is an identity matrix. Consequently,
\begin{equation}
    S(\rho_B^\mp) = \log_2(2) = 1.
\end{equation}
\
Conversely, the evaluation of the second term $- S(\rho_{\mathcal{N}_{\text{QS}}^{\ket{\mp}}}) \eqdef \text{Tr}\left[\rho_{\mathcal{N}_{\text{QS}}^{\ket{\mp}}}\log_2\rho_{\mathcal{N}_{\text{QS}}^{\ket{\mp}}}\right]$ in \eqref{eq:app:4} is more cumbersome since $\rho_{\mathcal{N}_{\text{QS}}^{\ket{\mp}}}$ is not diagonal. Indeed, to evaluate $\log_2\rho_{\mathcal{N}_{\text{QS}}^{\ket{\mp}}}$, it is convenient to determine the spectral decomposition of $\rho_{\mathcal{N}_{\text{QS}}^{\ket{\mp}}}$. In fact, by knowing the eigenvalues $\lambda_x$ of $\rho_{\mathcal{N}_{\text{QS}}^{\ket{\mp}}}$, the von Neumann entropy can be re-expressed as:
\begin{eqnarray}
	\label{eq:app:5}
	\nonumber -S(\rho_{\mathcal{N}_{\text{QS}}^{\ket{\mp}}}) &=& \text{Tr}\Bigl[\rho_{\mathcal{N}_{\text{QS}}^{\ket{\mp}}}\log_2\rho_{\mathcal{N}_{\text{QS}}^{\ket{\mp}}}\Bigr]\\
 &=& \sum_x \lambda_x^\mp \log_2 \lambda_x^\mp. 
\end{eqnarray}
Taking into account $p_0^{\mathcal{E}/\mathcal{D}} = 1 - p_X^{\mathcal{E}/\mathcal{D}} - p_Y^{\mathcal{E}/\mathcal{D}} - p_Z^{\mathcal{E}/\mathcal{D}}$, the eigenvalues $\lambda_x^\mp$ of $\rho_{\mathcal{N}_{\text{QS}}^{\ket{\mp}}}$ are given by:
\begin{align}
	\{ \lambda_x^-\} &= \left\{ 0, \frac{\eta_{XY}}{\eta}, \frac{\eta_{XZ}}{\eta}, \frac{\eta_{YZ}}{\eta} \right\}, \\
    \nonumber \{ \lambda_x^+\} &= \Bigl\{ \frac{\tilde{p}_X - \eta_{XY} - \eta_{XZ}}{1-\eta}, \frac{\tilde{p}_Y - \eta_{XY} - \eta_{YZ}}{1-\eta}, \\
     &\; \frac{\tilde{p}_Z - \eta_{XZ} - \eta_{YZ}}{1-\eta}, \frac{1 - \tilde{p} + \eta}{1-\eta} \Bigr\}.
\end{align}
Therefore, the corresponding von Neumann entropy reads:
\begin{align}
\nonumber -S(&\rho_{\mathcal{N}_{\text{QS}}^{\ket{-}}}) = \frac{1}{\eta}\Bigl( \eta_{XY}\log_2\eta_{XY} + \eta_{XZ}\log_2\eta_{XZ} \\
  \label{eq:VNEMinus} &+ \eta_{YZ}\log_2\eta_{YZ} \Bigr) - \log_2 \eta, \\
 \nonumber -S(&\rho_{\mathcal{N}_{\text{QS}}^{\ket{+}}}) = \frac{1}{1-\eta} \Bigl( (1 - \tilde{p} + \eta)\log_2(1 - \tilde{p} + \eta) \\
\nonumber &+ (\tilde{p}_X - \eta_{XY} - \eta_{XZ})\log_2(\tilde{p}_X - \eta_{XY} - \eta_{XZ}) \\
\nonumber &+ (\tilde{p}_Y - \eta_{XY} - \eta_{YZ})\log_2(\tilde{p}_Y - \eta_{XY} - \eta_{YZ}) \\
 \nonumber &+ (\tilde{p}_Z - \eta_{XZ} - \eta_{YZ})\log_2(\tilde{p}_Z - \eta_{XZ} - \eta_{YZ}) \Bigr) \\
  &-\log_2 (1-\eta). \label{eq:VNEPlus}
\end{align}
Finally, we evaluate the weighted coherent information of $\mathcal{N}_{\text{QS}}^{\ket{-}}$ by plugging into the obtained von Neumann entropies~\eqref{eq:VNEMinus} and~\eqref{eq:VNEPlus} and probabilities~\eqref{eq:app:pMinus} and~\eqref{eq:app:pPlus}:
\begin{align}
\nonumber p_- &I_c(\mathcal{N}_{\text{QS}}^{\ket{-}}) =  \eta(1 -  \log_2\eta) + \eta_{XY}\log_2\eta_{XY} \\
  &+ \eta_{XZ}\log_2\eta_{XZ} + \eta_{YZ}\log_2\eta_{YZ} \Bigr), \\
 \nonumber p_+ &I_c(\mathcal{N}_{\text{QS}}^{\ket{+}}) = (1-\eta)(1 - \log_2 (1-\eta)) \\
\nonumber &+ (\tilde{p}_X - \eta_{XY} - \eta_{XZ})\log_2(\tilde{p}_X - \eta_{XY} - \eta_{XZ}) \\
\nonumber &+ (\tilde{p}_Y - \eta_{XY} - \eta_{YZ})\log_2(\tilde{p}_Y - \eta_{XY} - \eta_{YZ}) \\
 &+ \nonumber (\tilde{p}_Z - \eta_{XZ} - \eta_{YZ})\log_2(\tilde{p}_Z - \eta_{XZ} - \eta_{YZ}) \\
 &+ (1 - \tilde{p} + \eta)\log_2(1 - \tilde{p} + \eta), \end{align}
from which the lower bound~\eqref{Eq:37} on the quantum path capacity for arbitrary Pauli channels follows.

\section{{Proof of Proposition~\ref{prop:4}}}
\label{App:4}

We start by borrowing the upper bound~\eqref{eq:app:11.1} on capacity of $\mathcal{N}_{\text{QS}}^{\ket{\mp}}(\cdot)$ from the relative entropy of entanglement of its Choi matrix:
\begin{equation}
	\label{eq:app:11:inf}
	\mathcal{Q}(\mathcal{N}_{\text{QS}}^{\ket{\mp}})\leq E_R(\rho_{\mathcal{N}_{\text{QS}}^{\ket{\mp}}}) \eqdef \inf_{\zeta_s^\mp} S(\rho_{\mathcal{N}_{\text{QS}}^{\ket{\mp}}}||\zeta_s^\mp),
\end{equation}
where $\zeta_s^\mp$ is an arbitrary separable state, and $S(\rho_{\mathcal{N}_{\text{QS}}^{\ket{\mp}}}||\zeta_s^\mp)$ is the \textit{relative entropy} of $\rho_{\mathcal{N}_{\text{QS}}^{\ket{\mp}}}$ with respect to it (see Def.~\ref{def:QRE}). Since $E_R(\rho_{\mathcal{N}_{\text{QS}}^{\ket{\mp}}})$ is defined as an infimum of $S(\rho_{\mathcal{N}_{\text{QS}}^{\ket{\mp}}}||\zeta_s^\mp)$ with respect to $\zeta_s^\mp$, it is upper-bounded as
\begin{equation}
	\label{eq:app.13}
	\mathcal{Q}(\mathcal{N}_{\text{QS}}^{\ket{\mp}})\leq E_R(\rho_{\mathcal{N}_{\text{QS}}^{\ket{\mp}}}) \leq S(\rho_{\mathcal{N}_{\text{QS}}^{\ket{\mp}}}||\zeta_s^\mp),
\end{equation}
for arbitrary $\zeta_s^\mp$. On the other hand,
\begin{align}
	\label{eq:app:12}
	\nonumber S(&\rho_{\mathcal{N}_{\text{QS}}^{\ket{\mp}}}||\zeta_s) \eqdef \text{Tr}\left[\rho_{\mathcal{N}_{\text{QS}}^{\ket{\mp}}}\left(\log_2\rho_{\mathcal{N}_{\text{QS}}^{\ket{+}}} - \log_2 \zeta_s^\mp\right)\right] \\
		\nonumber & = \text{Tr}\left[\rho_{\mathcal{N}_{\text{QS}}^{\ket{+}}}\log_2\rho_{\mathcal{N}_{\text{QS}}^{\ket{\mp}}}\right] - \text{Tr}\left[\rho_{\mathcal{N}_{\text{QS}}^{\ket{\mp}}}\log_2 \zeta_s^\mp\right] \\
  & \eqdef - S(\rho_{\mathcal{N}_{\text{QS}}^{\ket{\mp}}}) - \text{Tr}\left[\rho_{\mathcal{N}_{\text{QS}}^{\ket{\mp}}}\log_2 \zeta_s^\mp\right]. 
\end{align}
Let us choose the separable state $\zeta_s$ by reasoning as in \cite{PirLauOtt-17}, i.e.:
\begin{equation}
	\label{eq:app.14}
	\zeta_s^\mp= \frac{1}{2} \sum_{i=0}^1 \ket{i}\bra{i} \otimes \mathcal{N}_{\text{QS}}^{\ket{\mp}}(\ket{i}\bra{i}).
\end{equation}

By exploiting the Kraus decomposition of $\mathcal{N}_{\text{QS}}^{\ket{\mp}}(\cdot)$ with respect to the corresponding probability 
vectors~\eqref{eq:PauliGenCompMinus} and~\eqref{eq:PauliGenCompPlus}, after some algebraic manipulations and taking into account $p_0^{\mathcal{E}/\mathcal{D}} = 1 - p_X^{\mathcal{E}/\mathcal{D}} - p_Y^{\mathcal{E}/\mathcal{D}} - p_Z^{\mathcal{E}/\mathcal{D}}$, \eqref{eq:app.14} can be re-written as the following diagonal matrices:
\begin{align}
    \zeta_s^- &= \text{diag}\Bigl( \frac{\eta_{XY}}{2\eta}, \frac{\eta_{XZ}+\eta_{YZ}}{2\eta}, \frac{\eta_{XZ}+\eta_{YZ}}{2\eta}, \frac{\eta_{XY}}{2\eta}\Bigr), \\
    \nonumber \zeta_s^+ &= \text{diag}\Bigl( \frac{1-\tilde{p}_X - \tilde{p}_Y + \eta_{XY}}{2(1-\eta)}, \\
    \nonumber &\; \frac{\tilde{p}_X + \tilde{p}_Y - \eta_{XY} - \eta}{2(1-\eta)}, \frac{\tilde{p}_X + \tilde{p}_Y - \eta_{XY} - \eta}{2(1-\eta)}, \\ 
    &\; \frac{1-\tilde{p}_X - \tilde{p}_Y + \eta_{XY}}{2(1-\eta)} \Bigr).
\end{align}
While the von Neumann entropy $S(\rho_{\mathcal{N}_{\text{QS}}^{\ket{\mp}}})$ is known from~\eqref{eq:VNEMinus} and~\eqref{eq:VNEPlus}, the second term in~\eqref{eq:app:12} is less trivial. Denoting $\rho_{\mathcal{N}_{\text{QS}}^{\ket{\mp}}}\log_2 \zeta_s^\mp \eqdef \Xi^\mp$, after some algebraic manipulations, we obtain 
\begin{align}
    \nonumber \Xi^- &= \frac{\eta_{XY}}{\eta}\log_2\Bigl(\frac{\eta_{XY}}{2\eta}\Bigr)\rho_{\Phi^-} \\
    \nonumber &+ \frac{\eta_{YZ}}{\eta} \log_2 \Bigl( \frac{\eta_{XZ}+\eta_{YZ}}{2\eta} \Bigr) \rho_{\Psi^+} \\
    &+ \frac{\eta_{XZ}}{\eta} \log_2 \Bigl( \frac{\eta_{XZ}+\eta_{YZ}}{2\eta} \Bigr) \rho_{\Psi^-}, \\
    \nonumber \Xi^+ &= \frac{1-\tilde{p}_X - \tilde{p}_Y + \eta_{XY}}{1-\eta} \\
    \nonumber &\cdot \log_2 \Bigl( \frac{1-\tilde{p}_X - \tilde{p}_Y + \eta_{XY}}{2(1-\eta)} \Bigr) \rho_{\Phi^+} \\
    \nonumber &+ \frac{\tilde{p}_Z - \eta_{YZ} - \eta_{XZ}}{1-\eta} \\
    \nonumber &\cdot \log_2 \Bigl( \frac{1-\tilde{p}_X - \tilde{p}_Y + \eta_{XY}}{2(1-\eta)} \Bigr) (\rho_{\Phi^-} \! - \! \rho_{\Phi^+}), \\
    \nonumber &+ \frac{\tilde{p}_X - \eta_{XZ}}{1-\eta} \log_2 \Bigl( \frac{\tilde{p}_X + \tilde{p}_Y - \eta_{XY} - \eta}{2(1-\eta)}\Bigr) \rho_{\Psi^+} \\
    \nonumber &+ \frac{\tilde{p}_Y - \eta_{YZ}}{1-\eta} \log_2 \Bigl( \frac{\tilde{p}_X + \tilde{p}_Y - \eta_{XY} - \eta}{2(1-\eta)}\Bigr) \rho_{\Psi^-} \\
    &- \frac{\eta_{XY}}{1-\eta} \log_2 \Bigl( \frac{\tilde{p}_X + \tilde{p}_Y - \eta_{XY} - \eta}{2(1-\eta)}\Bigr) (\rho_{\Psi^+} \! + \! \rho_{\Psi^-}).
\end{align}
Taking into account that $\text{Tr}[\rho_{\Phi^\pm}] = \text{Tr}[\rho_{\Psi^\pm}] = 1$, we obtain:
\begin{align}
    \nonumber \text{Tr}[&\Xi^-] = \frac{1}{\eta}\Bigl( \eta_{XY}\log_2\eta_{XY} + (\eta_{YZ} + \eta_{XZ}) \\
    &\cdot \log_2 ( \eta_{YZ}+\eta_{XZ})\Bigr) - 1 - \log_2 \eta, \\
    \nonumber \text{Tr}[&\Xi^+] = \frac{1-\tilde{p}_X - \tilde{p}_Y + \eta_{XY}}{1-\eta} \\
    \nonumber &\cdot \log_2 ( 1-\tilde{p}_X - \tilde{p}_Y + \eta_{XY} ) \\
    \nonumber &+ \frac{\tilde{p}_X + \tilde{p}_Y - \eta_{XY} - \eta}{1-\eta} \\
   &\cdot  \log_2 ( \tilde{p}_X + \tilde{p}_Y - \eta_{XY} - \eta) - 1 - \log_2(1-\eta).
\end{align}
Finally, plugging in the obtained result to~\eqref{eq:app:12} and weighting it by the probabilities~\eqref{eq:app:pMinus} and~\eqref{eq:app:pPlus}, we find:
\begin{align}
	\nonumber p_- &S(\rho_{\mathcal{N}_{\text{QS}}^{\ket{-}}}||\zeta_s^\mp) = \eta + \eta_{YZ} \log_2 \eta_{YZ}  \\
 \nonumber &+ \eta_{XZ} \log_2 \eta_{XZ} - (\eta_{YZ} + \eta_{XZ}) \\
 &\cdot \log_2 ( \eta_{YZ}+\eta_{XZ}), \\
 \nonumber p_+ &S(\rho_{\mathcal{N}_{\text{QS}}^{\ket{+}}}||\zeta_s^\mp) = 1 - \eta \\
    \nonumber &+ (\tilde{p}_X - \eta_{XY} - \eta_{XZ})\log_2(\tilde{p}_X - \eta_{XY} - \eta_{XZ}) \\
\nonumber &+ (\tilde{p}_Y - \eta_{XY} - \eta_{YZ})\log_2(\tilde{p}_Y - \eta_{XY} - \eta_{YZ}) \\
\nonumber &+ (\tilde{p}_Z - \eta_{XZ} - \eta_{YZ})\log_2(\tilde{p}_Z - \eta_{XZ} - \eta_{YZ}) \\
 \nonumber &- (1-\tilde{p}_X - \tilde{p}_Y + \eta_{XY}) \\
 \nonumber &\cdot \log_2 ( 1-\tilde{p}_X - \tilde{p}_Y + \eta_{XY} ) \\
    \nonumber &- (\tilde{p}_X + \tilde{p}_Y - \eta_{XY} - \eta) \\
    \nonumber &\cdot \log_2 ( \tilde{p}_X + \tilde{p}_Y - \eta_{XY} - \eta) \\
       &+ (1 - \tilde{p} + \eta)\log_2(1 - \tilde{p} + \eta).
\end{align}
By summing both obtained terms up the upper bound~\eqref{Eq:38} on the quantum path's capacity for arbitrary Pauli channels follows.

\small
\bibliographystyle{IEEEtran}
\bibliography{main}

\begin{IEEEbiography}
[{\includegraphics[width=1in,height=1.25in,clip,keepaspectratio]{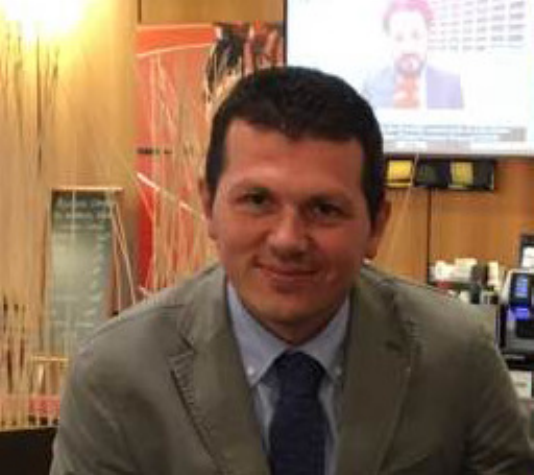}}]{Marcello Caleffi} (M'12, SM'16) is Associate Professor with the DIETI Department, University of Naples Federico II, where he co-lead the Quantum Internet research group. He is also with the National Laboratory of Multimedia Communications, National Inter-University Consortium for Telecommunications (CNIT). From 2010 to 2011, he was with the Broadband Wireless Networking Laboratory at Georgia Institute of Technology, as visiting researcher. In 2011, he was also with the NaNoNetworking Center in Catalunya (N3Cat) at the Universitat Politecnica de Catalunya (UPC), as visiting researcher. Since July 2018, he held the Italian national habilitation as \textit{Full Professor} in Telecommunications Engineering. His work appeared in several premier IEEE Transactions and Journals, and he received multiple awards, including \textit{best strategy}, \textit{most downloaded article}, and \textit{most cited article} awards. Currently, he serves as \textit{editor} for IEEE Trans. on Wireless Communications and IEEE Trans. on Quantum Engineering. Previously, he served as editor/associate technical editor for IEEE Communications Magazine and IEEE Communications Letters. He has served as Chair, TPC Chair, and TPC Member for several premier IEEE conferences. In 2017, he has been appointed as as \textit{distinguished lecturer} from the \textit{IEEE Computer Society} and he has been elected treasurer of the IEEE ComSoc/VT Italy Chapter. In 2019, he has been also appointed as member of the IEEE New Initiatives Committee from the \textit{IEEE Board of Directors}. In 2022, he has been awarded with the IEEE Communications Society ``\textit{Best Tutorial Paper Award}'' 2022 for the paper ``When Entanglement Meets Classical Communications: Quantum Teleportation for the Quantum Internet''.
\end{IEEEbiography}

\begin{IEEEbiography}
[{\includegraphics[width=1in,height=1.25in,clip,keepaspectratio]{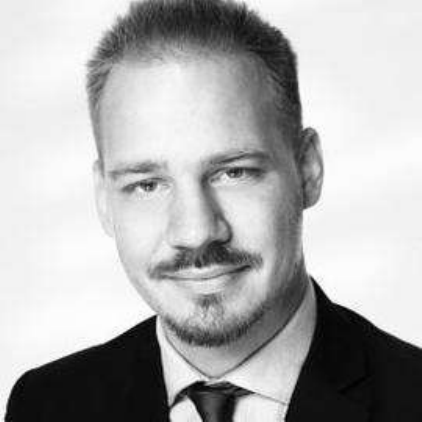}}]{Kyrylo Simonov} received the M.Sc. degree in physics in 2014 from the Taras Shevchenko National University of Kyiv (Ukraine) with a thesis on physics of DNA and the Ph.D. degree in physics in 2018 from the University of Vienna (Austria) with a thesis on quantum foundations. Since 2018 he worked at the Faculty of Mathematics of the University of Vienna (Austria) on mathematical foundations of quantum mechanics and applications of nonstandard analysis. His research interests include quantum information theory, quantum communications, quantum foundations, quantum thermodynamics, and mathematical foundations of quantum theory.
\end{IEEEbiography}

\begin{IEEEbiography}
[{\includegraphics[width=1in,height=1.25in,clip,keepaspectratio]{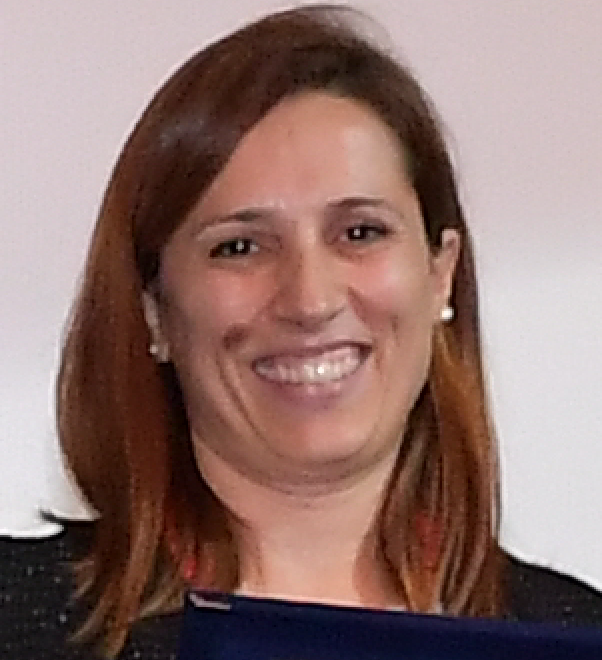}}]
{Angela Sara Cacciapuoti} (M'10, SM'16) is a professor at the University of Naples Federico II (Italy). Since July 2018 she held the national habilitation as “Full Professor” in Telecommunications Engineering. Her work has appeared in first tier IEEE journals and she has received different awards and recognition, including the \textit{``2022 IEEE ComSoc Best Tutorial Paper Award''} and \textit{``2021 N2Women: Stars in Networking and Communications''}. For the Quantum Internet topics, she is a \textit{IEEE ComSoc Distinguished Lecturer}, class of 2022-2023. Currently, Angela Sara serves as \textit{Area Editor} for IEEE Communications Letters, as \textit{Editor at Large} for IEEE Trans. on Communications and as \textit{Editor/Associate Editor} for the journals: IEEE Communications Surveys $\&$ Tutorials, IEEE Trans. on Wireless Communications, IEEE Trans. on Quantum Engineering, IEEE Network. She was the recipient of the \textit{2017 Exemplary Editor Award} of the IEEE Communications Letters. From 2020 to 2021, Angela Sara was the Vice-Chair of the IEEE ComSoc Women in Communications Engineering (WICE). Previously, she has been appointed as Publicity Chair of WICE. From 2016 to 2019 she has been an appointed member of the IEEE ComSoc Young Professionals Standing Committee. From 2017 to 2020, she has been the Treasurer of the IEEE Women in Engineering (WIE) Affinity Group of the IEEE Italy Section. Her current research interests are mainly in Quantum Communications, Quantum Networks and Quantum Information Processing.
\end{IEEEbiography}

\balance

\end{document}